\documentclass[twocolumn,showpacs,preprintnumbers,footinbib,prd,superscriptaddress,groupedaddress,10pt]{revtex4-1}

\usepackage[utf8]{inputenc}

\makeatletter
\def\l@subsubsection#1#2{}
\def\l@subsubsubsection#1#2{}
\makeatother

\usepackage{graphicx,amssymb,amsmath,amsthm,amsfonts}
\usepackage[usenames]{color}

\usepackage{notes2bib}

\usepackage{aas_macros}
\usepackage{bm}
\usepackage{dcolumn}
\usepackage{latexsym}
\usepackage{rotating}
\usepackage{longtable}

\usepackage{enumerate}
\usepackage{tensor,multirow}
\usepackage{url}
\usepackage[linktocpage]{hyperref}

\def\be{\begin{equation}}
\def\ee{\end{equation}}
\def\beq{\begin{eqnarray}}
\def\eeq{\end{eqnarray}}

\begin{document}

\title{Stirred and shaken: dynamical behavior of boson stars and dark matter cores}

\author{
Lorenzo Annulli,
Vitor Cardoso,
Rodrigo Vicente
}
\affiliation{${^1}$ Centro de Astrof\'{\i}sica e Gravita\c c\~ao - CENTRA, Departamento de F\'{\i}sica, Instituto
  Superior T\'ecnico - IST, Universidade de Lisboa - UL, Avenida Rovisco Pais 1, 1049-001 Lisboa, Portugal}

\begin{abstract} 
Bosonic fields can give rise to self-gravitating structures. These are interesting hypothetical new ``dark matter stars'' and good descriptions of dark matter haloes if the fields are very light. We study the dynamical response of Newtonian boson stars (NBS) when excited by external matter (stars, planets or black holes) in their vicinities. Our setup can describe the interaction between a massive black hole and the surrounding environment, shortly after the massive body has undergone a ``kick'', due to the collapse of baryonic matter at the galactic center, or dark matter depletion as a reaction to an inspiralling binary. We perform the first self-consistent calculation of dynamical friction acting on moving bodies in these backgrounds.
Binaries close to coalescence ``stir'' the NBS core, and backreaction affects gravitational waveforms at leading $-6PN$ order with respect to the dominant quadrupolar term; the coefficient is too small to allow detection by next-generation interferometers. We also show that the gravitational collapse to a supermassive black hole
at the center of a NBS is accompanied by only a small change in the surrounding core. The NBS eventually gets accreted, but for astrophysical parameters this occurs only after several Hubble times.
\end{abstract}

\maketitle

\noindent{\bf{\em Introduction.}}
The nature and properties of dark matter (DM) are arguably among the most important open issues in science. 
Interesting candidates for DM include light bosonic fields, of which the axion is a prototypical example~\cite{Peccei:1977hh}, or generalizations thereof such as
axion-like particles~\cite{Jaeckel:2010ni,Essig:2013lka}, ubiquitous in string-inspired scenarios~\cite{Arvanitaki:2009fg,Acharya:2015zfk}. 
Due to moduli compactifications, the spectrum of these particles could be populated uniformly down to the Hubble scale, $m_H\sim10^{-33}\,{\rm eV}$. 
The phenomenology of bosons with masses $\ll{\rm eV}$ can be dramatically different from that of weakly-interacting massive particles at the GeV-TeV scale. Ultralight bosons can form macroscopic Bose-Einstein condensates which provide a natural alternative to the standard structure formation through DM seeds and to the cold DM paradigm~\cite{Matos:1999et,Suarez:2013iw,Li:2013nal,Hui:2016ltb}. It has been recently recognized that ultralight boson fields with masses of the order of $10^{-22}\,{\rm eV}$ are a compelling candidate for cold DM~\cite{Robles:2012uy,Hui:2016ltb,Bar:2019bqz,Bar:2018acw,Desjacques:2019zhf}. A similar, albeit much wider, phenomenology arises in models of ultralight vector fields, such as dark photons, also a generic prediction of string theory~\cite{Goodsell:2009xc}. All experiments probing the interaction between DM and standard model
particles have so far been inconclusive. Given that DM interacts gravitationally,
and that practically all galaxies are permeated in this elusive matter, it is natural to study further its gravitational interaction. This prospect is made all the more attractive 
with the advent of gravitational-wave (GW) astronomy~\cite{Abbott:2016blz,LIGOScientific:2018mvr,Barack:2018yly}.

Light bosons can clump to form self-gravitating, stellar-mass or supermassive ``boson stars''~\cite{Kaup:1968zz,Ruffini:1969qy,Liebling:2012fv}.
The study of the dynamics of such objects is important for a number of reasons, ranging from stability to the way they interact with surrounding bodies (stars, BHs, etc)~\cite{Macedo:2013qea,Khlopov:1985}. 
It is crucial to understand scalar and GW emission, and possible constraints imposed via observations~\cite{Davoudiasl:2019nlo}. Along these lines, we have in mind the understanding of local changes in the density triggered by the presence of a massive BH or star, the drag exerted by the bosonic clump on stars moving within it, the flux of energy and momentum induced by coalescing binaries, etc.
These issues are relevant for large scale DM structures and GW physics~\cite{Eda:2013gg,Macedo:2013qea,Barausse:2014tra,Cardoso:2019rou,Kavanagh:2020cfn},
but also from the perspective of the interaction between DM stars and neutron stars or black holes (BHs)~\cite{Cardoso:2019rvt}.
These questions can also be of direct interest for theories with screening mechanisms, where new degrees of freedom -- usually scalars --
can be nonlinearly screened on some scales~\cite{Babichev:2013usa}. Such mechanisms give rise to nontrivial profiles for the new degrees of freedom,
for which many of the questions above apply.

We study the response of localized scalar configurations to bodies moving in their vicinities. 
The objects themselves -- Newtonian boson stars (NBSs) -- have been studied for decades~\cite{Kaup:1968zz,Ruffini:1969qy,Liebling:2012fv}. Despite this and the recent activity at the numerical relativity level~\cite{Cardoso:2016oxy,Helfer:2018vtq,Palenzuela:2017kcg,Sanchis-Gual:2019ljs,Bezares:2018qwa,Sanchis-Gual:2018oui,Widdicombe:2019woy}, their interaction with smaller objects (describing, for example, stars piercing through or orbiting such NBSs) has hardly been studied. The variety and disparity of scales in the problem makes it ill-suited for full-blown numerical techniques, but ideal for perturbation theory. Newtonian boson stars are structures made of complex scalar fields, which are a good description of dark matter haloes~\cite{Liebling:2012fv,Lee:1995af}. In the Newtonian limit, these objects are very similar to {\it oscillatons}~\cite{Guzman:2004wj}, which are made of real scalars (e.g. axion-like particles). The only difference concerns a small oscillating component of the gravitational potential. However, at the dynamical level, the response of both real and complex self-gravitating scalar structures is captured by our framework.

We use units where the speed of light, Newton's constant
and reduced Planck's constant are all set to unity,
$c=G=\hbar=1$.

\noindent{\bf{\em Setup: background.}}
We consider a general $U(1)$-invariant, self-interacting, complex scalar field $\Phi(x^\mu)$ minimally coupled to gravity, described by the action
\be
\mathcal{S}\equiv \int d^4x \sqrt{-g}\left(\frac{R}{16 \pi}-\frac{1}{2} g^{\mu\nu}\partial_{\mu}\Phi\partial_{\nu}\Phi^*-\frac{\mu^2}{2}|\Phi|^2\right)\,,\nonumber
\ee
where $R$ is the Ricci scalar of the spacetime metric $g_{\mu \nu}$, $g\equiv \det(g_{\mu \nu})$.
The scalar field stress-energy tensor $T^S_{\mu \nu }=\partial_{(\mu}\Phi^* \partial_{\nu)}\Phi-\frac{1}{2}g_{\mu \nu}\left[\partial_\alpha \Phi^* \partial^\alpha \Phi +\mu^2|\Phi|^2 \right]$.
For a spherically symmetric, stationary NBS
\begin{equation} 
\Phi=\Psi(r)e^{-i\Omega t}\,,\label{BKG_ansatz}
\end{equation}
where $\Psi(r)$ is a real function, regular everywhere. 
One can perform the Newtonian limit of the theory, with a spacetime ($|U(r)|\ll1$)
\be
ds^2=-\left(1+2U\right)dt^2+dr^2+r^2\left(d\theta^2+\sin^2\theta d\varphi^2\right)\,.
\ee
Assuming that the scalar $\Phi$ is non-relativistic, one finds the leading-order Schr\"{o}dinger-Poisson
equations of motion ($\nabla^2$ is the Laplacian operator)
\beq
\nabla^2 \Psi= 2 \mu \left(\mu U+\gamma\right) \Psi\,, \quad \nabla^2 U = 4\pi \mu^2 \Psi^2\,,\label{EOM_BS_radial}
\eeq
where non-relativistic regime implies that 
\be
\Omega = \mu-\gamma\,, \text{ with } 0<\gamma\ll\mu\,.
\ee
tu In this case, the energy $\Omega$ of the individual scalar \textit{particles} forming the NBS is approximately given by their rest-mass energy $\mu$. Remarkably, this system is left invariant under the transformation
\be
(\Psi,U,\gamma) \to \lambda^2 (\Psi, U, \gamma)\,,\quad  r \to r/\lambda\,.\label{eq:scaling}
\ee
These relations are crucial: once a NBS solution is found, all others can be obtained through a rescaling of that solution.
The numerical solution of system~\eqref{EOM_BS_radial}, with appropriate boundary conditions, is straightforward~\cite{Page:2003rd,Boskovic:2018rub,Kling:2017mif,Annulli:2020}.
Accurate and simple fits for the potential and scalar are shown in an upcoming work~\cite{Annulli:2020}.
At large distances, the scalar decays exponentially as $\Psi \sim e^{-\sqrt{2\mu\gamma}r}/r$, whereas the Newtonian potential falls off as $-M_{\rm NBS}/r$.
For fundamental NBSs one finds $\gamma\simeq0.162712 M_{\rm NBS}^2\mu^3$, where $M_{\rm NBS}$ is the NBS gravitational mass.
All the fundamental NBSs satisfy the scaling-invariant mass-radius relation $M_{\rm NBS}\mu\simeq 9.1/(R_{\rm NBS}\mu)$
where the NBS radius $R_{\rm NBS}$ is defined as the radius of the sphere enclosing $98\%$ of its mass~\cite{Liebling:2012fv,Boskovic:2018rub,Bar:2018acw}. We find the coupling $GM_{\rm NBS}\mu/(c\hbar)=7.5\times10^{-3}\left(M_{\rm NBS}/10^{10}M_\odot\right) \left(\mu/10^{-22}{\rm eV}\right)$.

\noindent{\bf{\em Setup: perturbations.}}
We are interested in the dynamical response to external perturbers, which disturb the spherically symmetric, stationary background above,
\begin{equation} 
\Phi=\left[\Psi_0(r)+\delta \Psi(t,r,\theta, \varphi)\right] e^{-i \Omega t}\,,\label{Perturbation}
\end{equation}
with the assumption $|\delta \Psi|\ll 1$, where $\Psi_0$ is the radial profile of the undisturbed NBS. The fluctuation $\delta \Psi$ induces a change $\delta T_{\mu \nu}^S$
in the stress-tensor which can be used to calculate physical quantities such as energy, linear and angular momenta radiated in a given process, obtained by computing the flux of certain currents,
\beq
&& \hspace{-0.2cm}\left(\dot{E}^{\rm rad},\dot{P}^{\rm rad}_i,\dot{L}^{\rm rad}_z\right)\nonumber\\
&& \hspace{-0.2cm}=\lim_{r\to \infty} r^2\int d\theta d\varphi \sin \theta\, \left(\delta T_{r \mu}^S \xi_t^\mu, \delta T_{r \mu}^S e_i^\mu\,, \delta T_{r \mu}^S \xi_\varphi^\mu \right)\,.
\eeq
Here, $\boldsymbol{\xi}_t= -\partial_t$, $\boldsymbol{\xi}_\varphi= \partial_\varphi$ are timelike and spacelike Killing vector fields of the background spacetime, respectively. The limit is being taken at fixed retarded time. The index $i=\{x,y,z\}$ and $\boldsymbol{e}_x$, $\boldsymbol{e}_y$, $\boldsymbol{e}_z$ are unit spacelike vectors in the $x$, $y$, $z$ directions.

Low-energy fluctuation modes, with energy $\omega^2<\left(\mu-\Omega\right)^2$,
are confined within the NBS. The interaction with an external perturber may deposit a considerable amount of energy and momentum in these modes.
Together with fluxes at infinity, these determine, for example, the energy loss of the perturber or the dynamical friction it is subjected to.
The deposition in the background object happens through excitation of its normal modes, which can be quantified 
once the dynamical equations for the perturbations have been established.
%
%

Non-relativistic perturbations of the form~\eqref{Perturbation} and corresponding gravitational potential fluctuation $\delta U$
satisfy the linearized system~\cite{Annulli:2020}
\beq
i \partial_{t} \delta \Psi&=&-\frac{1}{2 \mu}\nabla^2 \delta \Psi+ \left(\mu U_0 +\gamma\right) \delta \Psi+ \mu \Psi_0 \delta U\,,\label{Sourced_SP_System1}\\
\nabla^2 \delta U&=&4 \pi\left[ \mu^2 \Psi_0 \left(\delta \Psi+\delta \Psi^*\right)+P\right]\,,\label{Sourced_SP_System2}
\eeq
with the perturber treated as one or more pointlike sources, each of rest-mass $m_p$,
\be
P\equiv m_p \frac{\delta \left(r-r_p(t)\right)}{r^2} \frac{\delta\left(\theta-\theta_p(t)\right)}{\sin \theta} \delta\left(\varphi-\varphi_p(t)\right)\,,\label{source_BS}
\ee
with world line $x_p^\mu \equiv \left(t,r_p(t),\theta_p(t),\varphi_p(t)\right)$. 
We perform an expansion of $\delta\Psi,\,\delta U$ in spherical harmonics, and convert the equations to Fourier space with $\delta\Psi,\delta U\sim e^{\pm i\omega t}$. The end result is a set of coupled ordinary differential equations which can be robustly solved with standard techniques~\cite{Annulli:2020}.
Note that this system is invariant under the re-scaling $(U_0, \Psi_0,\gamma, \omega) \to \lambda^2 (U_0, \Psi_0, \gamma, \omega),\, r \to r/\lambda$.
Thus, one can map perturbations of a single NBS to the entire family of solutions.

\begin{table}[th] 
	\begin{tabular}{c||c}
		\hline
		\hline
		$l$ &  \multicolumn{1}{c}{$\omega^{(n)}_{\rm QNM}/(M_{\rm NBS}^2\mu^3)$} \\ 
		\hline
		\hline
		0 & $0.068\;\;\;  0.121\;\;\;     0.138\;\;\;    0.146\;\;\;    0.151\;\;\;  0.154\;\;\; 0.159$\\
		1 & $0.111\;\;\;     0.134\;\;\;     0.144\;\;\;    0.149\;\;\;    0.153\;\;\;  0.157\;\;\; 0.162$\\
		2 & $0.106\;\;\;     0.131\;\;\;     0.143\;\;\;    0.149\;\;\;    0.153\;\;\;  0.156\;\;\; 0.161$\\
		\hline
		\hline
	\end{tabular} 
	\caption{Normal frequencies of a NBS of mass $M_{\rm NBS}$ for the three lowest multipoles. Left to right are different overtone $n$ solutions.
	At large overtone number the modes cluster around $\gamma\simeq0.162712 M_{\rm NBS}^2\mu^3$. The first mode for $l=0$ agrees with that of Ref.~\cite{Guzman:2004wj} when properly normalized.}
	\label{table:QNM_BS_invariant}
\end{table}
The perturbative scheme requires that $|\delta \Psi|\ll 1$, which can always be enforced by making $m_p$ as small as necessary.
The background construction neglects higher-order PN contributions. A self-consistent perturbative expansion requires that such neglected terms $\sim U_0^2$ do not affect the dynamics $\sim \delta U$ of small fluctuations, imposing 
$m_p\gtrsim 10^4M_{\odot}\,\left(\frac{M_{\rm NBS}}{10^{10}M_\odot}\right)^3\left(\frac{\mu}{10^{-22}\,{\rm eV}}\right)^2$
which holds true for many systems of astrophysical interest (this constraint can be relaxed for purely dynamical quantities, the focus of this work).
Finally, the Newtonian, non-relativistic approximation requires the source to have a small frequency $\lesssim 10^{-7}\,\left(\mu/10^{-22}{\rm eV}\right)\,{\rm Hz}$.

Treating perturbers as pointlike is a successful and standard tool in BH perturbation theory~\cite{Zerilli:1971wd,Davis:1971gg,Barack:2018yvs}, in seismology~\cite{Ari} or in calculations of gravitational drag by fluids~\cite{Ostriker:1998fa,Vicente:2019ilr}. In this approximation one loses
small-scale information. For ultralight fields the smallest scale external to the pointlike particle is the Compton wavelength of the field,
which is much larger than the size of stars, planets or BHs. In other words, we do not expect to lose important details of the physics at play.

\noindent{\bf{\em Free oscillations.}}
The characteristic, non-relativistic oscillations of NBSs are regular solutions of the system \eqref{Sourced_SP_System1}-\eqref{Sourced_SP_System2}
satisfying Sommerfeld conditions at large distances. The first few characteristic frequencies are listed in Table~\ref{table:QNM_BS_invariant}.
They are all {\it normal mode} solutions, with real frequencies, confined within the NBS. The characteristic frequencies cluster around $\gamma$.
The fundamental monopolar mode in the Table agrees with one published result
after proper normalization~\cite{Guzman:2004wj}. Modes of relativistic stars have been considered in the literature~\cite{Yoshida:1994xi,Kojima:1991np,Macedo:2013jja,Macedo:2016wgh,GRITJHU}
and should smoothly go over to the numbers in Table~\ref{table:QNM_BS_invariant}.  Finally, we point out also that the stabilization of a newly formed boson star through the emission of scalar field was studied in Ref.\cite{Seidel1994}.

\noindent{\bf{\em An impurity at the center.}}
Static perturbations of NBSs are interesting in their own right. For perturbers far away from the NBS center, the induced tidal effects can dissipate energy
and lead to distinct signatures, both in GW signals and in the dynamics of objects close to such configurations~\cite{Mendes:2016vdr,Cardoso:2017cfl,Sennett:2017etc}.
The fluctuation in the density induced by a massive object sitting at the center is obtained in Ref.~\cite{Annulli:2020}.
The particle attracts scalar field towards the center, where the gravitational potential corresponds solely to that of the point-like mass. 
We find an insignificant change in the local DM mass density, $\delta \rho_M(0)/\rho_M(0)\sim 10\, m_p/M_{\rm NBS}$.

\noindent{\bf{\em A BH eating its host NBS.}}
On the other hand, studies of particle-like DM find that its density close to BHs increases significantly~\cite{Gondolo:1999ef,Sadeghian:2013laa}.
Such fact is in some tension with observations~\cite{Robles:2012uy}, but can be explained via scattering of DM by stars or BHs, or accretion by the central BH, induced by heating in its vicinities~\cite{Merritt:2002vj,Bertone:2005hw,Merritt:2003qk}. These results do not generalize to light scalars, at least when the configuration is spherically symmetric, since there are no stationary BH configurations with scalar ``hair''~\cite{Herdeiro:2015waa,Cardoso:2016ryw}. If a (non-spinning) BH forms at the center of a NBS, the scalar is accreted by the BH, thus our previous results cannot be extrapolated to this situation, and describe the system only at intermediate times. 
What {\it is} the lifetime of such a system? When $M_{\rm BH}\ll M_{\rm NBS}$, a perturbative calculation suffices. Consider a sphere of radius $r_+$ centered at the origin of the NBS. The NBS is stationary, and there is a flux of energy $\dot{E}_{\rm in}\approx 10^{-3} \mu^7 r_+^2 M_{\rm NBS}^5$ crossing such
a sphere inwards and outwards~\cite{Annulli:2020}. When a BH horizon exists at $r_+=2M_{\rm BH}$~\footnote{Actually, such a sphere should be placed outside the effective potential for wave propagation around BHs, but the distinction is not relevant here.}, a fraction of the incoming flux is absorbed by the BH. Low-frequency waves ($\mu M_{\rm BH}\ll 1$) are poorly absorbed~\cite{Unruh:1976fm}, and when $\omega \sim \mu$
\be
\dot{E}_{\rm abs}=32\pi\left(M_{\rm BH}\mu \right)^3\dot{E}_{\rm in}=\frac{16\pi}{125}\frac{M_{\rm BH}^5}{M_{\rm NBS}^5}\left(M_{\rm NBS}\mu \right)^{10}\,.\nonumber
\ee
We have tested the above physics with a series of toy models, including a massive, non self-gravitating, scalar confined in a spherical cavity
with a small BH at the center~\cite{Annulli:2020}. The models conform to the physics just outlined.

With $\dot{E}_{\rm abs}=\dot{M}_{\rm BH}$ and fixed NBS mass, one finds the timescale
\beq
\tau\sim 
%
10^{33}\,{\rm yr}\,\frac{M_{\rm NBS}}{10^{10}M_{\odot}}\left(\frac{\chi}{10^4}\right)^4\left(\frac{0.01}{M_{\rm NBS}\mu}\right)^{10}\,,
\eeq
where $\chi\equiv M_{\rm NBS}/M_{\rm BH}$. In other words, the timescale for the BH to increase substantially its mass -- a conservative indicative of the lifetime of the entire NBS -- is larger than a Hubble timescale for realistic parameters. This timescale is the result of forcing the BH with the nearly monochromatic field of the NBS. When the NBS material is nearly exhausted, a new timescale is relevant, that of the quasinormal modes of a BH surrounded by a massive scalar $\sim M_{\rm BH}(M_{\rm BH}\mu)^{-6}<\tau$, still typically larger than a Hubble time.  
When rotation is included, the entire setup may become even more stable; superradiance from the BH might even support nearly stationary, but non spherically-symmetric, configurations~\cite{Herdeiro:2014goa,Brito:2015oca}.

\noindent{\bf{\em Plunges into NBSs.}}
%
\begin{figure}
\includegraphics[width=8cm,keepaspectratio]{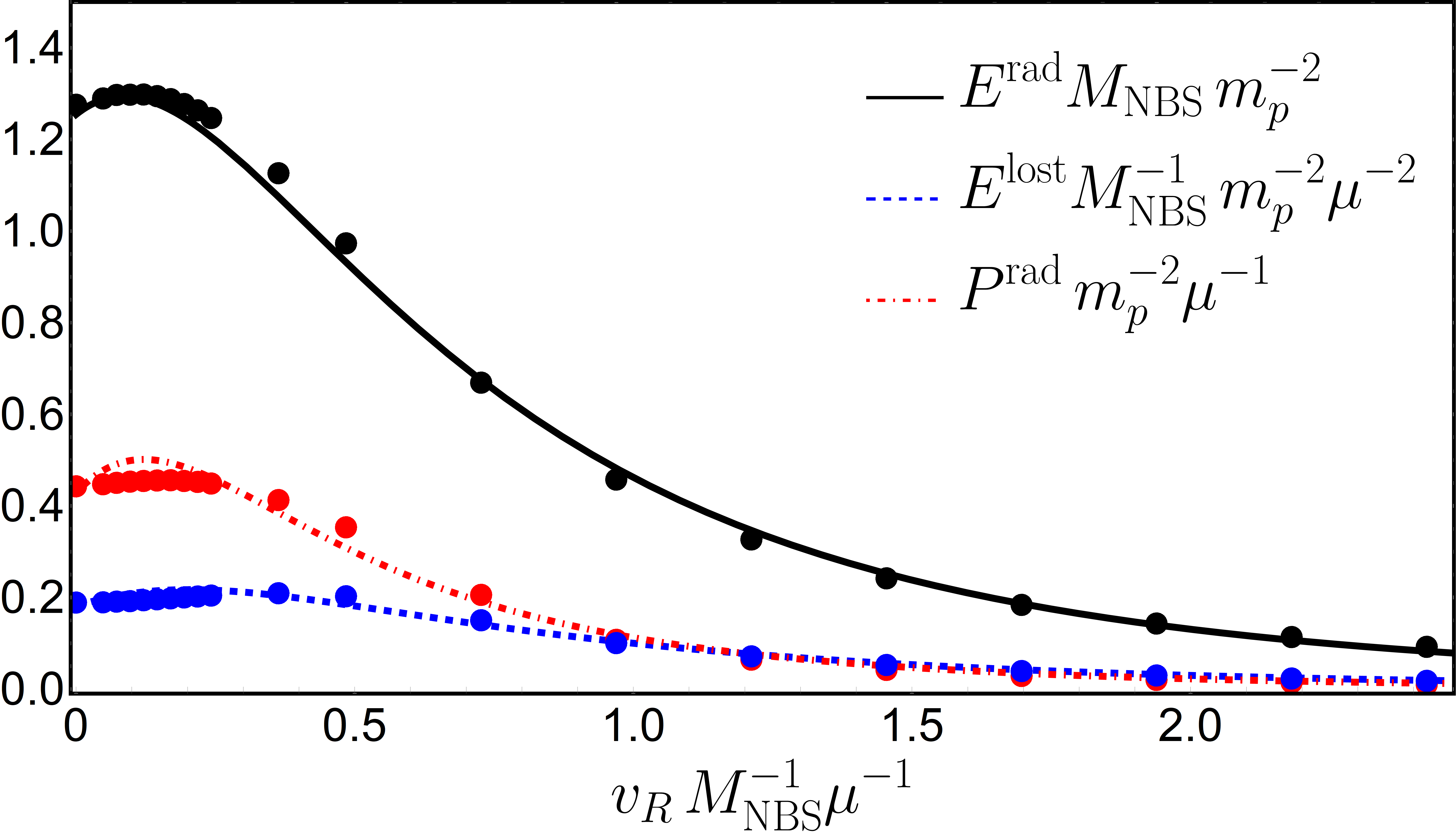} 
\caption{Total and kinetic energy, and linear momentum emitted when an object of mass $m_p$ plunges through an NBS, as a function of the initial perturber velocity. The dots correspond to the numerical data, the solid curves are best-fits, such as Eq.~\eqref{eq:fit_BS_Ekin_plunge_gravityon}.}
	\label{fig:fitsBS}
\end{figure}
Baryonic matter tends to slowly accumulate near the center of a DM structure, where it may eventually collapse to a massive BH. Gravitational collapse can impart a recoil velocity $v_{\rm recoil}$ to the BH, of the order of $300\,{\rm Km/sec}$~\cite{1973ApJ...183..657B}. This will cause a BH to oscillate within the DM halo, on timescales of
$\sim  10^6 \,{\rm yr} \sqrt{10^3M_{\odot}{\rm pc}^{-3}/\rho}$,
and with an amplitude $\sim 69\, {\rm pc}\,\frac{v_{\rm recoil}}{300 \, {\rm Km/s}} \sqrt{10^3M_{\odot}{\rm pc}^{-3}/\rho}$, well within our non-relativistic approximation.
To model these processes, consider first a particle plunging head-on into a NBS, with velocity $\boldsymbol{v}=-v_R \boldsymbol{e}_z$ (with $v>0$) at the NBS surface. The particle crosses the NBS exactly once, and for small $v_R$ the process is always well-described by a Newtonian, non-relativistic approximation.

The motion triggers a flux of scalar particles at infinity. The radiation and momentum spectrum $E^{\rm rad},\,P^{\rm rad}$ are shown in Fig.~\ref{fig:fitsBS} for different $v_R$.
The energy lost by the crossing particle, $E^{\rm lost}$, is also shown and is only a fraction of $E^{\rm rad}$, since the scalars have nonzero rest-mass.
We find the following good description of our results 
\be 
E^{\rm lost}=7\,m_p^2 M_{\rm NBS}^{21/4}\mu^{25/4}\frac{e^{-3.54M_{\rm NBS}\mu/\left(v_R+0.63M_{\rm NBS}\mu\right)}}{\left(v_R+0.63M_{\rm NBS}\mu\right)^{17/4}}\,,\label{eq:fit_BS_Ekin_plunge_gravityon}
\ee
accurate to within $5\%$ of error for~$0\lesssim v_R\lesssim 2.5M_{\rm NBS}\mu$. This interval spans over non-relativistic astrophysical relevant velocities ({\it e.g} $0\lesssim v_R[{\rm km/s}]\lesssim 6000$) for the DM core of the Milky Way.

The momentum lost by the plunging object is given by $P^{\rm lost}=E^{\rm lost}/v_R$, and also gives us the dynamical friction exerted on the traveling body, $dP/dt \sim P^{\rm lost}v/(2R_{\rm NBS})$. Previous estimates, using an infinite, non self-gravitating scalar as an approximation~\cite{Hui:2016ltb,Lancaster:2019mde}, are plagued by near and far-distance divergences. By contrast, our results are finite and regular in the small-velocity limit, and are the first self-consistent calculation of the energy and momentum released when a massive body travels through a self-gravitating medium. Including the self-gravity of the scalar is a crucial ingredient for a correct calculation of dynamical friction (details to appear in Ref.~\cite{Annulli:2020}). At low velocities, our results predict a friction force one order of magnitude smaller than estimated without all the ingredients.

We can also calculate the relaxation timescale of a BH undergoing bound motion inside a NBS. For small-amplitude motion, we find that the amplitude $\mathcal{A}$
of the motion decays exponentially in tie as a consequence of dynamical friction, $\mathcal{A}=\mathcal{A}_0\, e^{-t/\tau_{\rm s}}$
with the timescale
\be
\tau_{s}
\sim 10^{10} {\rm yr} \left(\frac{10^{-22}\, {\rm eV}}{\mu}\right)^2\left(\frac{10^5 M_ \odot}{m_p}\right)\left(\frac{0.01}{M_ {\rm NBS}\mu}\right)\,. 
\ee
This result described the timescale for a kicked BH (or star) to settle down at the center of an halo purely due to the dynamical friction caused by dark matter. 
We find that, excluding all but the interaction with the scalar, BHs of mass $m_p\gtrsim 10^5 M_\odot$ settle down in a timescale smaller than the Hubble time.
This result can be compared with the timescale of damping due to dynamical friction caused by stars in the galactic core, estimated to be $\sim 0.1\,\tau_s$ if we take
the mass in the galactic core $M_{\rm c}=M_{\rm NBS}$~\cite{Gualandris:2007nm}. Given the associated uncertainties, these results imply that DM needs to be taken into account when studying such processes.

\noindent{\bf{\em Low-energy binaries within NBSs.}}
Consider now compact binaries within a DM core. These could be two BH binaries formed via core collapse or two neutron star binaries somewhere close to the galactic center.
These systems have been observed recently via electromagnetic counterparts to GWs~\cite{Graham:2020gwr}.
Binaries, either at an early or late stage in their life,
stir the scalar field and produce disturbances in its profile. 
At high ($\omega_{\rm orb} \gg \gamma, \mu U_0$), but still non-relativistic frequencies ($\omega_{\rm orb} \ll \mu$) 
 equations~\eqref{Sourced_SP_System1} and~\eqref{Sourced_SP_System2} imply that $|\Psi_0 \delta \Psi|\ll |\delta U|$. Thus, equation~\eqref{Sourced_SP_System2} reduces simply to 
$\nabla^2 \delta U=4 \pi P$. In this regime, we find the following analytic solution,
%
\begin{align}
&\dot{E}^{\rm lost}\simeq 0.28\, \pi^{3} \left(\mu m_p\right)^2 \left(\mu M_{\rm NBS}\right)^4 \sum_{m=1}^{+\infty}\left[1+(-1)^m\right]^2 \nonumber\\ &\times \left(\frac{Y_m^m\left(\frac{\pi}{2},0\right)}{\Gamma\left(m+\frac{3}{2}\right)} \frac{m^{\left(\frac{m}{2}-\frac{1}{4}\right)}(M\omega_{\rm orb})^{\frac{m}{3}}}{2^{\left(\frac{7}{4}+\frac{m}{2}\right)}(\omega_{\rm orb}/\mu)^{\left(\frac{1}{4}+\frac{m}{2}\right)}}\right)^2\Theta\left[m\omega_{\rm orb}-\gamma\right]\,.
\end{align}
Here, $M=2m_p$ for equal-mass binaries. The substitution $\left(1+(-1)^m\right)\to 1$ describes a single particle of mass $m_p$ revolving around a BH of mass $M_{\rm BH}=M\gg m_p$.
The analytic expression above agrees with a full numerical solution within $\sim 2\%$ in the entire regime of validity.
At fixed orbital frequency the flux converges exponentially in $l$. Emission starts at $\omega=\gamma\ll \mu$, and the coupling between gravity and the scalar implies
that large frequency sources radiate less. The energy emitted in scalar waves by equal-mass binaries, with respect to their own quadrupole GW flux, $\dot{E}^{\rm GW}=(32/5) m_p^2 M^{-2}\left(M\omega_{\rm orb}\right)^{10/3}$, reads as
\be
\frac{\dot{E}^{\rm lost}}{\dot{E}^{\rm GW}}\sim 10^{-5}\left[\frac{M_{\rm NBS}\mu}{0.01}\right]^4\left[\frac{\mu}{10^{-22}\,{\rm eV}}\right]^{\frac{9}{2}}\left[\frac{T_{\rm orb}}{16\,{\rm yrs}}\right]^{\frac{9}{2}}\,,\nonumber
\ee
showing promising numbers, for low frequency emitters.
\noindent{\bf{\em High-energy binaries within NBSs.}}
Rapidly moving binaries, such as those suitable for LIGO or LISA sources, do not fit into a non-relativistic regime. The relevant description of these systems for frequencies $\omega_{\rm orb}\gg \mu$ is~\cite{Annulli:2020}
\be
\nabla^2\delta U= 4\pi P\,,\quad-\partial^2_t \delta \Phi +\nabla^2 \delta \Phi =2 \mu^2 \Phi\, \delta U\,.\label{eq:high_binary}
\ee
We find the following simple result for the flux in the dominant $l=m$ modes, 
\beq
\dot{E}^{\rm rad}&\simeq&\dot{E}^{\rm lost}=128 \pi^{3}(\mu^2 m_p \Psi_0(0))^2\left(1+(-1)^m\right)^2  \nonumber\\ 
&&\times\sum_{m=1}^{+\infty}\left(\frac{Y_m^m(\pi/2,0)}{\Gamma(m+3/2)} \frac{m^{m-1}(M\omega_{\rm orb})^{m/3}}{2^{m+1}\,\omega_{\rm orb}}\right)^2.\label{eq:energy_loss_high_binaries}
\eeq
Again, $\left(1+(-1)^m\right)\to 1$ describes a single particle of mass $m_p$ revolving around a massive object.

To understand the impact of such energy loss on GW signals, we add to the quadrupole formula the energy flux~\eqref{eq:energy_loss_high_binaries}, 
and compute the correction to the GW phase in the stationary approximation~\cite{Flanagan:1997sx}. In Fourier space, we decompose the phase of the GW signal 
$\tilde{h}(f)={\cal A}e^{i\Upsilon(f)}$ as $\Upsilon(f) =\Upsilon^{(0)}[1+\delta_{\Upsilon}]$,
where $\Upsilon^{(0)}=3/128 ({\cal M}\pi f)^{-5/3}$ represents the leading term of the 
phase's post-Newtonian expansion, and $f=\omega_{\rm orb}/\pi$. We find
\be
\delta_{\Upsilon}=\frac{16\mu^4\Psi_0^2}{51\pi^3f^4}\sim 10^{-24}\left[\frac{\mu}{10^{-22}\,{\rm eV}}\right]^4\left[\frac{10^{-4}}{f}\right]^4\left[\frac{M_{\rm NBS}\mu}{0.01}\right]^4\nonumber
\ee
for the scalar contribution in equal-mass binaries. Such a correction corresponds to a $-6 PN$ order contribution, and we have normalized it against possible values for LISA and galactic cores. Even though it is a very low and negative PN correction, its magnitude is so small as to make its detection via GWs seem hopeless~\cite{Barausse:2014tra,Cardoso:2019rou}.

\noindent{\bf{\em Discussion.}}
This work shows how self-gravitating NBSs respond to time-varying, localized matter fluctuations.
These are structures that behave classically: they are composed of $N ~\sim 10^{100}\left(10^{-22}{\rm eV}/\mu\right)^2$ particles; a binary of two supermassive BHs in the last stages of coalescence emits more than $10^{60}$ particles. Our results show unique features of bosonic ultralight structures. For example, they are not easily depleted by binaries. Even a supermassive BH binary close to coalescence would need a Hubble time or more to completely deplete the scalar in a sphere of ten-wavelength radius around the binary. In other words, the perturbative framework is consistent and robust.

This study should be extended to eccentric motion, or to self-gravitating vectorial configurations or even other nonlinearly interacting scalars~\cite{Coleman:1985ki}.
Our results should also be a useful benchmark for numerical relativity simulations involving boson stars in the extreme mass ratio regime, when and if the field is able to accommodate such challenging setups.

\section*{Acknowledgements}
%
We are indebted to the Theory Institute at CERN and to Waseda University for warm hospitality while this work was being completed.
We thank Ana Sousa Carvalho for advice and support, and Emanuele Berti for useful comments and suggestions.
V.~C.\ acknowledges financial support provided under the European Union's H2020 ERC 
Consolidator Grant ``Matter and strong-field gravity: New frontiers in Einstein's 
theory'' grant agreement no. MaGRaTh--646597.
R.V.\ was  supported by the FCT PhD scholarship SFRH/BD/128834/2017.
L.A. acknowledges financial support provided by Funda\c{c}ao para a Ci\^{e}ncia e a Tecnologia Grant number PD/BD/128232/2016 awarded in the framework of the Doctoral Programme IDPASC-Portugal.
This project has received funding from the European Union's Horizon 2020 research and innovation 
programme under the Marie Sklodowska-Curie grant agreement No 690904.
We thank FCT for financial support through Project~No.~UIDB/00099/2020.
We acknowledge financial support provided by FCT/Portugal through grant PTDC/MAT-APL/30043/2017.
The authors would like to acknowledge networking support by the GWverse COST Action 
CA16104, ``Black holes, gravitational waves and fundamental physics.''
%

\bibliographystyle{apsrev4}
\bibliography{References}

\begin{thebibliography}{74}%
\makeatletter
\providecommand \@ifxundefined [1]{%
 \@ifx{#1\undefined}
}%
\providecommand \@ifnum [1]{%
 \ifnum #1\expandafter \@firstoftwo
 \else \expandafter \@secondoftwo
 \fi
}%
\providecommand \@ifx [1]{%
 \ifx #1\expandafter \@firstoftwo
 \else \expandafter \@secondoftwo
 \fi
}%
\providecommand \natexlab [1]{#1}%
\providecommand \enquote  [1]{``#1''}%
\providecommand \bibnamefont  [1]{#1}%
\providecommand \bibfnamefont [1]{#1}%
\providecommand \citenamefont [1]{#1}%
\providecommand \href@noop [0]{\@secondoftwo}%
\providecommand \href [0]{\begingroup \@sanitize@url \@href}%
\providecommand \@href[1]{\@@startlink{#1}\@@href}%
\providecommand \@@href[1]{\endgroup#1\@@endlink}%
\providecommand \@sanitize@url [0]{\catcode `\\12\catcode `\$12\catcode
  `\&12\catcode `\#12\catcode `\^12\catcode `\_12\catcode `\%12\relax}%
\providecommand \@@startlink[1]{}%
\providecommand \@@endlink[0]{}%
\providecommand \url  [0]{\begingroup\@sanitize@url \@url }%
\providecommand \@url [1]{\endgroup\@href {#1}{\urlprefix }}%
\providecommand \urlprefix  [0]{URL }%
\providecommand \Eprint [0]{\href }%
\providecommand \doibase [0]{http://dx.doi.org/}%
\providecommand \selectlanguage [0]{\@gobble}%
\providecommand \bibinfo  [0]{\@secondoftwo}%
\providecommand \bibfield  [0]{\@secondoftwo}%
\providecommand \translation [1]{[#1]}%
\providecommand \BibitemOpen [0]{}%
\providecommand \bibitemStop [0]{}%
\providecommand \bibitemNoStop [0]{.\EOS\space}%
\providecommand \EOS [0]{\spacefactor3000\relax}%
\providecommand \BibitemShut  [1]{\csname bibitem#1\endcsname}%
\let\auto@bib@innerbib\@empty
\bibitem [{\citenamefont {Peccei} and \citenamefont
  {Quinn}(1977)}]{Peccei:1977hh}%
  \BibitemOpen
  \bibfield  {author} {\bibinfo {author} {\bibfnamefont {R.D.} \bibnamefont
  {Peccei}} and \bibinfo {author} {\bibfnamefont {H.R.} \bibnamefont {Quinn}},
  }\href {\doibase 10.1103/PhysRevLett.38.1440} {\bibfield  {journal} {\bibinfo
   {journal} {\emph {Phys. Rev. Lett.}} }\textbf {\bibinfo {volume} {38}},
  \bibinfo {pages} {1440} (\bibinfo {year} {1977})}\BibitemShut {NoStop}%
\bibitem [{\citenamefont {Jaeckel} and \citenamefont
  {Ringwald}(2010)}]{Jaeckel:2010ni}%
  \BibitemOpen
  \bibfield  {author} {\bibinfo {author} {\bibfnamefont {J.}~\bibnamefont
  {Jaeckel}} and \bibinfo {author} {\bibfnamefont {A.}~\bibnamefont
  {Ringwald}}, }\href {\doibase 10.1146/annurev.nucl.012809.104433} {\bibfield
  {journal} {\bibinfo  {journal} {\emph {Ann. Rev. Nucl. Part. Sci.}} }\textbf
  {\bibinfo {volume} {60}}, \bibinfo {pages} {405} (\bibinfo {year} {2010})},
  \Eprint {http://arxiv.org/abs/1002.0329} {arXiv:1002.0329}\BibitemShut
  {NoStop}%
\bibitem [{\citenamefont {Essig} \emph {et~al.}(2013)}]{Essig:2013lka}%
  \BibitemOpen
  \bibfield  {author} {\bibinfo {author} {\bibfnamefont {R.}~\bibnamefont
  {Essig}} \emph {et~al.}, }in \href
  {http://inspirehep.net/record/1263039/files/arXiv:1311.0029.pdf} {\emph
  {\bibinfo {booktitle} {{Community Summer Study 2013: Snowmass on the
  Mississippi (CSS2013) Minneapolis, MN, USA, July 29-August 6, 2013}}}}
  (\bibinfo {year} {2013}) \Eprint {http://arxiv.org/abs/1311.0029}
  {arXiv:1311.0029}\BibitemShut {NoStop}%
\bibitem [{\citenamefont {Arvanitaki} \emph {et~al.}(2010)\citenamefont
  {Arvanitaki}, \citenamefont {Dimopoulos}, \citenamefont {Dubovsky},
  \citenamefont {Kaloper}, and \citenamefont
  {March-Russell}}]{Arvanitaki:2009fg}%
  \BibitemOpen
  \bibfield  {author} {\bibinfo {author} {\bibfnamefont {A.}~\bibnamefont
  {Arvanitaki}}, \bibinfo {author} {\bibfnamefont {S.}~\bibnamefont
  {Dimopoulos}}, \bibinfo {author} {\bibfnamefont {S.}~\bibnamefont
  {Dubovsky}}, \bibinfo {author} {\bibfnamefont {N.}~\bibnamefont {Kaloper}},
  and \bibinfo {author} {\bibfnamefont {J.}~\bibnamefont {March-Russell}},
  }\href {\doibase 10.1103/PhysRevD.81.123530} {\bibfield  {journal} {\bibinfo
  {journal} {\emph {Phys.Rev.}} }\textbf {\bibinfo {volume} {D81}}, \bibinfo
  {pages} {123530} (\bibinfo {year} {2010})}, \Eprint
  {http://arxiv.org/abs/0905.4720} {arXiv:0905.4720}\BibitemShut {NoStop}%
\bibitem [{\citenamefont {Acharya} and \citenamefont
  {Pongkitivanichkul}(2016)}]{Acharya:2015zfk}%
  \BibitemOpen
  \bibfield  {author} {\bibinfo {author} {\bibfnamefont {B.S.} \bibnamefont
  {Acharya}} and \bibinfo {author} {\bibfnamefont {C.}~\bibnamefont
  {Pongkitivanichkul}}, }\href {\doibase 10.1007/JHEP04(2016)009} {\bibfield
  {journal} {\bibinfo  {journal} {\emph {JHEP}} }\textbf {\bibinfo {volume}
  {04}}, \bibinfo {pages} {009} (\bibinfo {year} {2016})}, \Eprint
  {http://arxiv.org/abs/1512.07907} {arXiv:1512.07907}\BibitemShut {NoStop}%
\bibitem [{\citenamefont {Matos} \emph {et~al.}(2000)\citenamefont {Matos},
  \citenamefont {Guzman}, and \citenamefont {Urena-Lopez}}]{Matos:1999et}%
  \BibitemOpen
  \bibfield  {author} {\bibinfo {author} {\bibfnamefont {T.}~\bibnamefont
  {Matos}}, \bibinfo {author} {\bibfnamefont {F.S.} \bibnamefont {Guzman}},
  and \bibinfo {author} {\bibfnamefont {L.}~\bibnamefont {Urena-Lopez}}, }\href
  {\doibase 10.1088/0264-9381/17/7/309} {\bibfield  {journal} {\bibinfo
  {journal} {\emph {Class. Quant. Grav.}} }\textbf {\bibinfo {volume} {17}},
  \bibinfo {pages} {1707} (\bibinfo {year} {2000})}, \Eprint
  {http://arxiv.org/abs/astro-ph/9908152} {arXiv:astro-ph/9908152}\BibitemShut
  {NoStop}%
\bibitem [{\citenamefont {Suarez} \emph {et~al.}(2014)\citenamefont {Suarez},
  \citenamefont {Robles}, and \citenamefont {Matos}}]{Suarez:2013iw}%
  \BibitemOpen
  \bibfield  {author} {\bibinfo {author} {\bibfnamefont {A.}~\bibnamefont
  {Suarez}}, \bibinfo {author} {\bibfnamefont {V.H.} \bibnamefont {Robles}},
  and \bibinfo {author} {\bibfnamefont {T.}~\bibnamefont {Matos}}, }\bibfield
  {booktitle} {\emph {\bibinfo {booktitle} {{Proceedings of 4th International
  Meeting on Gravitation and Cosmology (MGC 4)}}}, }\href {\doibase
  10.1007/978-3-319-02063-1_9} {\bibfield  {journal} {\bibinfo  {journal}
  {\emph {Astrophys. Space Sci. Proc.}} }\textbf {\bibinfo {volume} {38}},
  \bibinfo {pages} {107} (\bibinfo {year} {2014})}, \Eprint
  {http://arxiv.org/abs/1302.0903} {arXiv:1302.0903}\BibitemShut {NoStop}%
\bibitem [{\citenamefont {Li} \emph {et~al.}(2014)\citenamefont {Li},
  \citenamefont {Rindler-Daller}, and \citenamefont {Shapiro}}]{Li:2013nal}%
  \BibitemOpen
  \bibfield  {author} {\bibinfo {author} {\bibfnamefont {B.}~\bibnamefont
  {Li}}, \bibinfo {author} {\bibfnamefont {T.}~\bibnamefont {Rindler-Daller}},
  and \bibinfo {author} {\bibfnamefont {P.R.} \bibnamefont {Shapiro}}, }\href
  {\doibase 10.1103/PhysRevD.89.083536} {\bibfield  {journal} {\bibinfo
  {journal} {\emph {Phys. Rev.}} }\textbf {\bibinfo {volume} {D89}}, \bibinfo
  {pages} {083536} (\bibinfo {year} {2014})}, \Eprint
  {http://arxiv.org/abs/1310.6061} {arXiv:1310.6061}\BibitemShut {NoStop}%
\bibitem [{\citenamefont {Hui} \emph {et~al.}(2017)\citenamefont {Hui},
  \citenamefont {Ostriker}, \citenamefont {Tremaine}, and \citenamefont
  {Witten}}]{Hui:2016ltb}%
  \BibitemOpen
  \bibfield  {author} {\bibinfo {author} {\bibfnamefont {L.}~\bibnamefont
  {Hui}}, \bibinfo {author} {\bibfnamefont {J.P.} \bibnamefont {Ostriker}},
  \bibinfo {author} {\bibfnamefont {S.}~\bibnamefont {Tremaine}},  and \bibinfo
  {author} {\bibfnamefont {E.}~\bibnamefont {Witten}}, }\href {\doibase
  10.1103/PhysRevD.95.043541} {\bibfield  {journal} {\bibinfo  {journal} {\emph
  {Phys. Rev.}} }\textbf {\bibinfo {volume} {D95}}, \bibinfo {pages} {043541}
  (\bibinfo {year} {2017})}, \Eprint {http://arxiv.org/abs/1610.08297}
  {arXiv:1610.08297}\BibitemShut {NoStop}%
\bibitem [{\citenamefont {Robles} and \citenamefont
  {Matos}(2012)}]{Robles:2012uy}%
  \BibitemOpen
  \bibfield  {author} {\bibinfo {author} {\bibfnamefont {V.H.} \bibnamefont
  {Robles}} and \bibinfo {author} {\bibfnamefont {T.}~\bibnamefont {Matos}},
  }\href {\doibase 10.1111/j.1365-2966.2012.20603.x} {\bibfield  {journal}
  {\bibinfo  {journal} {\emph {Mon. Not. Roy. Astron. Soc.}} }\textbf {\bibinfo
  {volume} {422}}, \bibinfo {pages} {282} (\bibinfo {year} {2012})}, \Eprint
  {http://arxiv.org/abs/1201.3032} {arXiv:1201.3032}\BibitemShut {NoStop}%
\bibitem [{\citenamefont {Bar} \emph {et~al.}(2019)\citenamefont {Bar},
  \citenamefont {Blum}, \citenamefont {Eby}, and \citenamefont
  {Sato}}]{Bar:2019bqz}%
  \BibitemOpen
  \bibfield  {author} {\bibinfo {author} {\bibfnamefont {N.}~\bibnamefont
  {Bar}}, \bibinfo {author} {\bibfnamefont {K.}~\bibnamefont {Blum}}, \bibinfo
  {author} {\bibfnamefont {J.}~\bibnamefont {Eby}},  and \bibinfo {author}
  {\bibfnamefont {R.}~\bibnamefont {Sato}}, }\href {\doibase
  10.1103/PhysRevD.99.103020} {\bibfield  {journal} {\bibinfo  {journal} {\emph
  {Phys. Rev.}} }\textbf {\bibinfo {volume} {D99}}, \bibinfo {pages} {103020}
  (\bibinfo {year} {2019})}, \Eprint {http://arxiv.org/abs/1903.03402}
  {arXiv:1903.03402}\BibitemShut {NoStop}%
\bibitem [{\citenamefont {Bar} \emph {et~al.}(2018)\citenamefont {Bar},
  \citenamefont {Blas}, \citenamefont {Blum}, and \citenamefont
  {Sibiryakov}}]{Bar:2018acw}%
  \BibitemOpen
  \bibfield  {author} {\bibinfo {author} {\bibfnamefont {N.}~\bibnamefont
  {Bar}}, \bibinfo {author} {\bibfnamefont {D.}~\bibnamefont {Blas}}, \bibinfo
  {author} {\bibfnamefont {K.}~\bibnamefont {Blum}},  and \bibinfo {author}
  {\bibfnamefont {S.}~\bibnamefont {Sibiryakov}}, }\href {\doibase
  10.1103/PhysRevD.98.083027} {\bibfield  {journal} {\bibinfo  {journal} {\emph
  {Phys. Rev.}} }\textbf {\bibinfo {volume} {D98}}, \bibinfo {pages} {083027}
  (\bibinfo {year} {2018})}, \Eprint {http://arxiv.org/abs/1805.00122}
  {arXiv:1805.00122}\BibitemShut {NoStop}%
\bibitem [{\citenamefont {Desjacques} and \citenamefont
  {Nusser}(2019)}]{Desjacques:2019zhf}%
  \BibitemOpen
  \bibfield  {author} {\bibinfo {author} {\bibfnamefont {V.}~\bibnamefont
  {Desjacques}} and \bibinfo {author} {\bibfnamefont {A.}~\bibnamefont
  {Nusser}}, }\href {\doibase 10.1093/mnras/stz1978} {\bibfield  {journal}
  {\bibinfo  {journal} {\emph {Mon. Not. Roy. Astron. Soc.}} }\textbf {\bibinfo
  {volume} {488}}, \bibinfo {pages} {4497} (\bibinfo {year} {2019})}, \Eprint
  {http://arxiv.org/abs/1905.03450} {arXiv:1905.03450}\BibitemShut {NoStop}%
\bibitem [{\citenamefont {Goodsell} \emph {et~al.}(2009)\citenamefont
  {Goodsell}, \citenamefont {Jaeckel}, \citenamefont {Redondo}, and
  \citenamefont {Ringwald}}]{Goodsell:2009xc}%
  \BibitemOpen
  \bibfield  {author} {\bibinfo {author} {\bibfnamefont {M.}~\bibnamefont
  {Goodsell}}, \bibinfo {author} {\bibfnamefont {J.}~\bibnamefont {Jaeckel}},
  \bibinfo {author} {\bibfnamefont {J.}~\bibnamefont {Redondo}},  and \bibinfo
  {author} {\bibfnamefont {A.}~\bibnamefont {Ringwald}}, }\href {\doibase
  10.1088/1126-6708/2009/11/027} {\bibfield  {journal} {\bibinfo  {journal}
  {\emph {JHEP}} }\textbf {\bibinfo {volume} {11}}, \bibinfo {pages} {027}
  (\bibinfo {year} {2009})}, \Eprint {http://arxiv.org/abs/0909.0515}
  {arXiv:0909.0515}\BibitemShut {NoStop}%
\bibitem [{\citenamefont {Abbott} \emph {et~al.}(2016)}]{Abbott:2016blz}%
  \BibitemOpen
  \bibfield  {author} {\bibinfo {author} {\bibfnamefont {B.}~\bibnamefont
  {Abbott}} \emph {et~al.} (\bibinfo {collaboration} {LIGO Scientific, Virgo}),
  }\href {\doibase 10.1103/PhysRevLett.116.061102} {\bibfield  {journal}
  {\bibinfo  {journal} {\emph {Phys. Rev. Lett.}} }\textbf {\bibinfo {volume}
  {116}}, \bibinfo {pages} {061102} (\bibinfo {year} {2016})}, \Eprint
  {http://arxiv.org/abs/1602.03837} {arXiv:1602.03837}\BibitemShut {NoStop}%
\bibitem [{\citenamefont {Abbott} \emph
  {et~al.}(2019)}]{LIGOScientific:2018mvr}%
  \BibitemOpen
  \bibfield  {author} {\bibinfo {author} {\bibfnamefont {B.}~\bibnamefont
  {Abbott}} \emph {et~al.} (\bibinfo {collaboration} {LIGO Scientific, Virgo}),
  }\href {\doibase 10.1103/PhysRevX.9.031040} {\bibfield  {journal} {\bibinfo
  {journal} {\emph {Phys. Rev. X}} }\textbf {\bibinfo {volume} {9}}, \bibinfo
  {pages} {031040} (\bibinfo {year} {2019})}, \Eprint
  {http://arxiv.org/abs/1811.12907} {arXiv:1811.12907}\BibitemShut {NoStop}%
\bibitem [{\citenamefont {Barack} \emph {et~al.}(2019)}]{Barack:2018yly}%
  \BibitemOpen
  \bibfield  {author} {\bibinfo {author} {\bibfnamefont {L.}~\bibnamefont
  {Barack}} \emph {et~al.}, }\href {\doibase 10.1088/1361-6382/ab0587}
  {\bibfield  {journal} {\bibinfo  {journal} {\emph {Class.\ Quant.\ Grav.}}
  }\textbf {\bibinfo {volume} {36}}, \bibinfo {pages} {143001} (\bibinfo {year}
  {2019})}, \Eprint {http://arxiv.org/abs/1806.05195}
  {arXiv:1806.05195}\BibitemShut {NoStop}%
\bibitem [{\citenamefont {Kaup}(1968)}]{Kaup:1968zz}%
  \BibitemOpen
  \bibfield  {author} {\bibinfo {author} {\bibfnamefont {D.J.} \bibnamefont
  {Kaup}}, }\href {\doibase 10.1103/PhysRev.172.1331} {\bibfield  {journal}
  {\bibinfo  {journal} {\emph {Phys.Rev.}} }\textbf {\bibinfo {volume} {172}},
  \bibinfo {pages} {1331} (\bibinfo {year} {1968})}\BibitemShut {NoStop}%
\bibitem [{\citenamefont {Ruffini} and \citenamefont
  {Bonazzola}(1969)}]{Ruffini:1969qy}%
  \BibitemOpen
  \bibfield  {author} {\bibinfo {author} {\bibfnamefont {R.}~\bibnamefont
  {Ruffini}} and \bibinfo {author} {\bibfnamefont {S.}~\bibnamefont
  {Bonazzola}}, }\href {\doibase 10.1103/PhysRev.187.1767} {\bibfield
  {journal} {\bibinfo  {journal} {\emph {Phys.Rev.}} }\textbf {\bibinfo
  {volume} {187}}, \bibinfo {pages} {1767} (\bibinfo {year}
  {1969})}\BibitemShut {NoStop}%
\bibitem [{\citenamefont {Liebling} and \citenamefont
  {Palenzuela}(2012)}]{Liebling:2012fv}%
  \BibitemOpen
  \bibfield  {author} {\bibinfo {author} {\bibfnamefont {S.L.} \bibnamefont
  {Liebling}} and \bibinfo {author} {\bibfnamefont {C.}~\bibnamefont
  {Palenzuela}}, }\href {\doibase 10.12942/lrr-2012-6,
  10.1007/s41114-017-0007-y} {\bibfield  {journal} {\bibinfo  {journal} {\emph
  {Living Rev. Rel.}} }\textbf {\bibinfo {volume} {15}}, \bibinfo {pages} {6}
  (\bibinfo {year} {2012})}, \bibinfo {note} {[Living Rev.
  Rel.20,no.1,5(2017)]}, \Eprint {http://arxiv.org/abs/1202.5809}
  {arXiv:1202.5809}\BibitemShut {NoStop}%
\bibitem [{\citenamefont {Macedo} \emph
  {et~al.}(2013{\natexlab{a}})\citenamefont {Macedo}, \citenamefont {Pani},
  \citenamefont {Cardoso}, and \citenamefont {Crispino}}]{Macedo:2013qea}%
  \BibitemOpen
  \bibfield  {author} {\bibinfo {author} {\bibfnamefont {C.F.B.} \bibnamefont
  {Macedo}}, \bibinfo {author} {\bibfnamefont {P.}~\bibnamefont {Pani}},
  \bibinfo {author} {\bibfnamefont {V.}~\bibnamefont {Cardoso}},  and \bibinfo
  {author} {\bibfnamefont {L.C.B.} \bibnamefont {Crispino}}, }\href {\doibase
  10.1088/0004-637X/774/1/48} {\bibfield  {journal} {\bibinfo  {journal} {\emph
  {Astrophys. J.}} }\textbf {\bibinfo {volume} {774}}, \bibinfo {pages} {48}
  (\bibinfo {year} {2013}{\natexlab{a}})}, \Eprint
  {http://arxiv.org/abs/1302.2646} {arXiv:1302.2646}\BibitemShut {NoStop}%
\bibitem [{\citenamefont {{Khlopov}} \emph {et~al.}(1985)\citenamefont
  {{Khlopov}}, \citenamefont {{Malomed}}, and \citenamefont
  {{Zeldovich}}}]{Khlopov:1985}%
  \BibitemOpen
  \bibfield  {author} {\bibinfo {author} {\bibfnamefont {M.I.} \bibnamefont
  {{Khlopov}}}, \bibinfo {author} {\bibfnamefont {B.A.} \bibnamefont
  {{Malomed}}},  and \bibinfo {author} {\bibfnamefont {I.B.} \bibnamefont
  {{Zeldovich}}}, }\href {\doibase 10.1093/mnras/215.4.575} {\bibfield
  {journal} {\bibinfo  {journal} {\emph {Mon. Not. Roy. Astron. Soc.}} }\textbf
  {\bibinfo {volume} {215}}, \bibinfo {pages} {575} (\bibinfo {year}
  {1985})}\BibitemShut {NoStop}%
\bibitem [{\citenamefont {Davoudiasl} and \citenamefont
  {Denton}(2019)}]{Davoudiasl:2019nlo}%
  \BibitemOpen
  \bibfield  {author} {\bibinfo {author} {\bibfnamefont {H.}~\bibnamefont
  {Davoudiasl}} and \bibinfo {author} {\bibfnamefont {P.B.} \bibnamefont
  {Denton}}, }\href {\doibase 10.1103/PhysRevLett.123.021102} {\bibfield
  {journal} {\bibinfo  {journal} {\emph {Phys. Rev. Lett.}} }\textbf {\bibinfo
  {volume} {123}}, \bibinfo {pages} {021102} (\bibinfo {year} {2019})}, \Eprint
  {http://arxiv.org/abs/1904.09242} {arXiv:1904.09242}\BibitemShut {NoStop}%
\bibitem [{\citenamefont {Eda} \emph {et~al.}(2013)\citenamefont {Eda},
  \citenamefont {Itoh}, \citenamefont {Kuroyanagi}, and \citenamefont
  {Silk}}]{Eda:2013gg}%
  \BibitemOpen
  \bibfield  {author} {\bibinfo {author} {\bibfnamefont {K.}~\bibnamefont
  {Eda}}, \bibinfo {author} {\bibfnamefont {Y.}~\bibnamefont {Itoh}}, \bibinfo
  {author} {\bibfnamefont {S.}~\bibnamefont {Kuroyanagi}},  and \bibinfo
  {author} {\bibfnamefont {J.}~\bibnamefont {Silk}}, }\href {\doibase
  10.1103/PhysRevLett.110.221101} {\bibfield  {journal} {\bibinfo  {journal}
  {\emph {Phys.Rev.Lett.}} }\textbf {\bibinfo {volume} {110}}, \bibinfo {pages}
  {221101} (\bibinfo {year} {2013})}, \Eprint {http://arxiv.org/abs/1301.5971}
  {arXiv:1301.5971}\BibitemShut {NoStop}%
\bibitem [{\citenamefont {Barausse} \emph {et~al.}(2014)\citenamefont
  {Barausse}, \citenamefont {Cardoso}, and \citenamefont
  {Pani}}]{Barausse:2014tra}%
  \BibitemOpen
  \bibfield  {author} {\bibinfo {author} {\bibfnamefont {E.}~\bibnamefont
  {Barausse}}, \bibinfo {author} {\bibfnamefont {V.}~\bibnamefont {Cardoso}},
  and \bibinfo {author} {\bibfnamefont {P.}~\bibnamefont {Pani}}, }\href
  {\doibase 10.1103/PhysRevD.89.104059} {\bibfield  {journal} {\bibinfo
  {journal} {\emph {Phys. Rev.}} }\textbf {\bibinfo {volume} {D89}}, \bibinfo
  {pages} {104059} (\bibinfo {year} {2014})}, \Eprint
  {http://arxiv.org/abs/1404.7149} {arXiv:1404.7149}\BibitemShut {NoStop}%
\bibitem [{\citenamefont {Cardoso} and \citenamefont
  {Maselli}(2019)}]{Cardoso:2019rou}%
  \BibitemOpen
  \bibfield  {author} {\bibinfo {author} {\bibfnamefont {V.}~\bibnamefont
  {Cardoso}} and \bibinfo {author} {\bibfnamefont {A.}~\bibnamefont {Maselli}},
  }\href@noop {} {  (\bibinfo {year} {2019})}, \Eprint
  {http://arxiv.org/abs/1909.05870} {arXiv:1909.05870}\BibitemShut {NoStop}%
\bibitem [{\citenamefont {Kavanagh} \emph {et~al.}(2020)\citenamefont
  {Kavanagh}, \citenamefont {Nichols}, \citenamefont {Bertone}, and
  \citenamefont {Gaggero}}]{Kavanagh:2020cfn}%
  \BibitemOpen
  \bibfield  {author} {\bibinfo {author} {\bibfnamefont {B.J.} \bibnamefont
  {Kavanagh}}, \bibinfo {author} {\bibfnamefont {D.A.} \bibnamefont {Nichols}},
  \bibinfo {author} {\bibfnamefont {G.}~\bibnamefont {Bertone}},  and \bibinfo
  {author} {\bibfnamefont {D.}~\bibnamefont {Gaggero}}, }\href@noop {} {
  (\bibinfo {year} {2020})}, \Eprint {http://arxiv.org/abs/2002.12811}
  {arXiv:2002.12811}\BibitemShut {NoStop}%
\bibitem [{\citenamefont {Cardoso} and \citenamefont
  {Pani}(2019)}]{Cardoso:2019rvt}%
  \BibitemOpen
  \bibfield  {author} {\bibinfo {author} {\bibfnamefont {V.}~\bibnamefont
  {Cardoso}} and \bibinfo {author} {\bibfnamefont {P.}~\bibnamefont {Pani}},
  }\href {\doibase 10.1007/s41114-019-0020-4} {\bibfield  {journal} {\bibinfo
  {journal} {\emph {Living Rev. Rel.}} }\textbf {\bibinfo {volume} {22}},
  \bibinfo {pages} {4} (\bibinfo {year} {2019})}, \Eprint
  {http://arxiv.org/abs/1904.05363} {arXiv:1904.05363}\BibitemShut {NoStop}%
\bibitem [{\citenamefont {Babichev} and \citenamefont
  {Deffayet}(2013)}]{Babichev:2013usa}%
  \BibitemOpen
  \bibfield  {author} {\bibinfo {author} {\bibfnamefont {E.}~\bibnamefont
  {Babichev}} and \bibinfo {author} {\bibfnamefont {C.}~\bibnamefont
  {Deffayet}}, }\href {\doibase 10.1088/0264-9381/30/18/184001} {\bibfield
  {journal} {\bibinfo  {journal} {\emph {Class.\ Quant.\ Grav.}} }\textbf
  {\bibinfo {volume} {30}}, \bibinfo {pages} {184001} (\bibinfo {year}
  {2013})}, \Eprint {http://arxiv.org/abs/1304.7240}
  {arXiv:1304.7240}\BibitemShut {NoStop}%
\bibitem [{\citenamefont {Cardoso} \emph {et~al.}(2016)\citenamefont {Cardoso},
  \citenamefont {Hopper}, \citenamefont {Macedo}, \citenamefont {Palenzuela},
  and \citenamefont {Pani}}]{Cardoso:2016oxy}%
  \BibitemOpen
  \bibfield  {author} {\bibinfo {author} {\bibfnamefont {V.}~\bibnamefont
  {Cardoso}}, \bibinfo {author} {\bibfnamefont {S.}~\bibnamefont {Hopper}},
  \bibinfo {author} {\bibfnamefont {C.F.B.} \bibnamefont {Macedo}}, \bibinfo
  {author} {\bibfnamefont {C.}~\bibnamefont {Palenzuela}},  and \bibinfo
  {author} {\bibfnamefont {P.}~\bibnamefont {Pani}}, }\href {\doibase
  10.1103/PhysRevD.94.084031} {\bibfield  {journal} {\bibinfo  {journal} {\emph
  {Phys. Rev. D}} }\textbf {\bibinfo {volume} {94}}, \bibinfo {pages} {084031}
  (\bibinfo {year} {2016})}, \Eprint {http://arxiv.org/abs/1608.08637}
  {arXiv:1608.08637}\BibitemShut {NoStop}%
\bibitem [{\citenamefont {Helfer} \emph {et~al.}(2019)\citenamefont {Helfer},
  \citenamefont {Lim}, \citenamefont {Garcia}, and \citenamefont
  {Amin}}]{Helfer:2018vtq}%
  \BibitemOpen
  \bibfield  {author} {\bibinfo {author} {\bibfnamefont {T.}~\bibnamefont
  {Helfer}}, \bibinfo {author} {\bibfnamefont {E.A.} \bibnamefont {Lim}},
  \bibinfo {author} {\bibfnamefont {M.A.} \bibnamefont {Garcia}},  and \bibinfo
  {author} {\bibfnamefont {M.A.} \bibnamefont {Amin}}, }\href {\doibase
  10.1103/PhysRevD.99.044046} {\bibfield  {journal} {\bibinfo  {journal} {\emph
  {Phys. Rev. D}} }\textbf {\bibinfo {volume} {99}}, \bibinfo {pages} {044046}
  (\bibinfo {year} {2019})}, \Eprint {http://arxiv.org/abs/1802.06733}
  {arXiv:1802.06733}\BibitemShut {NoStop}%
\bibitem [{\citenamefont {Palenzuela} \emph {et~al.}(2017)\citenamefont
  {Palenzuela}, \citenamefont {Pani}, \citenamefont {Bezares}, \citenamefont
  {Cardoso}, \citenamefont {Lehner}, and \citenamefont
  {Liebling}}]{Palenzuela:2017kcg}%
  \BibitemOpen
  \bibfield  {author} {\bibinfo {author} {\bibfnamefont {C.}~\bibnamefont
  {Palenzuela}}, \bibinfo {author} {\bibfnamefont {P.}~\bibnamefont {Pani}},
  \bibinfo {author} {\bibfnamefont {M.}~\bibnamefont {Bezares}}, \bibinfo
  {author} {\bibfnamefont {V.}~\bibnamefont {Cardoso}}, \bibinfo {author}
  {\bibfnamefont {L.}~\bibnamefont {Lehner}},  and \bibinfo {author}
  {\bibfnamefont {S.}~\bibnamefont {Liebling}}, }\href {\doibase
  10.1103/PhysRevD.96.104058} {\bibfield  {journal} {\bibinfo  {journal} {\emph
  {Phys. Rev.}} }\textbf {\bibinfo {volume} {D96}}, \bibinfo {pages} {104058}
  (\bibinfo {year} {2017})}, \Eprint {http://arxiv.org/abs/1710.09432}
  {arXiv:1710.09432}\BibitemShut {NoStop}%
\bibitem [{\citenamefont {Sanchis-Gual} \emph
  {et~al.}(2019{\natexlab{a}})\citenamefont {Sanchis-Gual}, \citenamefont
  {Di~Giovanni}, \citenamefont {Zilhão}, \citenamefont {Herdeiro},
  \citenamefont {Cerdá-Durán}, \citenamefont {Font}, and \citenamefont
  {Radu}}]{Sanchis-Gual:2019ljs}%
  \BibitemOpen
  \bibfield  {author} {\bibinfo {author} {\bibfnamefont {N.}~\bibnamefont
  {Sanchis-Gual}}, \bibinfo {author} {\bibfnamefont {F.}~\bibnamefont
  {Di~Giovanni}}, \bibinfo {author} {\bibfnamefont {M.}~\bibnamefont
  {Zilhão}}, \bibinfo {author} {\bibfnamefont {C.}~\bibnamefont {Herdeiro}},
  \bibinfo {author} {\bibfnamefont {P.}~\bibnamefont {Cerdá-Durán}}, \bibinfo
  {author} {\bibfnamefont {J.}~\bibnamefont {Font}},  and \bibinfo {author}
  {\bibfnamefont {E.}~\bibnamefont {Radu}}, }\href {\doibase
  10.1103/PhysRevLett.123.221101} {\bibfield  {journal} {\bibinfo  {journal}
  {\emph {Phys. Rev. Lett.}} }\textbf {\bibinfo {volume} {123}}, \bibinfo
  {pages} {221101} (\bibinfo {year} {2019}{\natexlab{a}})}, \Eprint
  {http://arxiv.org/abs/1907.12565} {arXiv:1907.12565}\BibitemShut {NoStop}%
\bibitem [{\citenamefont {Bezares} and \citenamefont
  {Palenzuela}(2018)}]{Bezares:2018qwa}%
  \BibitemOpen
  \bibfield  {author} {\bibinfo {author} {\bibfnamefont {M.}~\bibnamefont
  {Bezares}} and \bibinfo {author} {\bibfnamefont {C.}~\bibnamefont
  {Palenzuela}}, }\href {\doibase 10.1088/1361-6382/aae87c} {\bibfield
  {journal} {\bibinfo  {journal} {\emph {Class. Quant. Grav.}} }\textbf
  {\bibinfo {volume} {35}}, \bibinfo {pages} {234002} (\bibinfo {year}
  {2018})}, \Eprint {http://arxiv.org/abs/1808.10732}
  {arXiv:1808.10732}\BibitemShut {NoStop}%
\bibitem [{\citenamefont {Sanchis-Gual} \emph
  {et~al.}(2019{\natexlab{b}})\citenamefont {Sanchis-Gual}, \citenamefont
  {Herdeiro}, \citenamefont {Font}, \citenamefont {Radu}, and \citenamefont
  {Di~Giovanni}}]{Sanchis-Gual:2018oui}%
  \BibitemOpen
  \bibfield  {author} {\bibinfo {author} {\bibfnamefont {N.}~\bibnamefont
  {Sanchis-Gual}}, \bibinfo {author} {\bibfnamefont {C.}~\bibnamefont
  {Herdeiro}}, \bibinfo {author} {\bibfnamefont {J.A.} \bibnamefont {Font}},
  \bibinfo {author} {\bibfnamefont {E.}~\bibnamefont {Radu}},  and \bibinfo
  {author} {\bibfnamefont {F.}~\bibnamefont {Di~Giovanni}}, }\href {\doibase
  10.1103/PhysRevD.99.024017} {\bibfield  {journal} {\bibinfo  {journal} {\emph
  {Phys. Rev. D}} }\textbf {\bibinfo {volume} {99}}, \bibinfo {pages} {024017}
  (\bibinfo {year} {2019}{\natexlab{b}})}, \Eprint
  {http://arxiv.org/abs/1806.07779} {arXiv:1806.07779}\BibitemShut {NoStop}%
\bibitem [{\citenamefont {Widdicombe} \emph {et~al.}(2020)\citenamefont
  {Widdicombe}, \citenamefont {Helfer}, and \citenamefont
  {Lim}}]{Widdicombe:2019woy}%
  \BibitemOpen
  \bibfield  {author} {\bibinfo {author} {\bibfnamefont {J.Y.} \bibnamefont
  {Widdicombe}}, \bibinfo {author} {\bibfnamefont {T.}~\bibnamefont {Helfer}},
  and \bibinfo {author} {\bibfnamefont {E.A.} \bibnamefont {Lim}}, }\href
  {\doibase 10.1088/1475-7516/2020/01/027} {\bibfield  {journal} {\bibinfo
  {journal} {\emph {JCAP}} }\textbf {\bibinfo {volume} {01}}, \bibinfo {pages}
  {027} (\bibinfo {year} {2020})}, \Eprint {http://arxiv.org/abs/1910.01950}
  {arXiv:1910.01950}\BibitemShut {NoStop}%
\bibitem [{\citenamefont {Lee} and \citenamefont {Koh}(1996)}]{Lee:1995af}%
  \BibitemOpen
  \bibfield  {author} {\bibinfo {author} {\bibfnamefont {J.w.} \bibnamefont
  {Lee}} and \bibinfo {author} {\bibfnamefont {I.g.} \bibnamefont {Koh}},
  }\href {\doibase 10.1103/PhysRevD.53.2236} {\bibfield  {journal} {\bibinfo
  {journal} {\emph {Phys. Rev. D}} }\textbf {\bibinfo {volume} {53}}, \bibinfo
  {pages} {2236} (\bibinfo {year} {1996})}, \Eprint
  {http://arxiv.org/abs/hep-ph/9507385} {arXiv:hep-ph/9507385}\BibitemShut
  {NoStop}%
\bibitem [{\citenamefont {Page}(2004)}]{Page:2003rd}%
  \BibitemOpen
  \bibfield  {author} {\bibinfo {author} {\bibfnamefont {D.N.} \bibnamefont
  {Page}}, }\href {\doibase 10.1103/PhysRevD.70.023002} {\bibfield  {journal}
  {\bibinfo  {journal} {\emph {Phys. Rev. D}} }\textbf {\bibinfo {volume}
  {70}}, \bibinfo {pages} {023002} (\bibinfo {year} {2004})}, \Eprint
  {http://arxiv.org/abs/gr-qc/0310006} {arXiv:gr-qc/0310006}\BibitemShut
  {NoStop}%
\bibitem [{\citenamefont {Boskovic} \emph {et~al.}(2018)\citenamefont
  {Boskovic}, \citenamefont {Duque}, \citenamefont {Ferreira}, \citenamefont
  {Miguel}, and \citenamefont {Cardoso}}]{Boskovic:2018rub}%
  \BibitemOpen
  \bibfield  {author} {\bibinfo {author} {\bibfnamefont {M.}~\bibnamefont
  {Boskovic}}, \bibinfo {author} {\bibfnamefont {F.}~\bibnamefont {Duque}},
  \bibinfo {author} {\bibfnamefont {M.C.} \bibnamefont {Ferreira}}, \bibinfo
  {author} {\bibfnamefont {F.S.} \bibnamefont {Miguel}},  and \bibinfo {author}
  {\bibfnamefont {V.}~\bibnamefont {Cardoso}}, }\href {\doibase
  10.1103/PhysRevD.98.024037} {\bibfield  {journal} {\bibinfo  {journal} {\emph
  {Phys.\ Rev.\ D}} }\textbf {\bibinfo {volume} {98}}, \bibinfo {pages}
  {024037} (\bibinfo {year} {2018})}, \Eprint {http://arxiv.org/abs/1806.07331}
  {arXiv:1806.07331}\BibitemShut {NoStop}%
\bibitem [{\citenamefont {Kling} and \citenamefont
  {Rajaraman}(2017)}]{Kling:2017mif}%
  \BibitemOpen
  \bibfield  {author} {\bibinfo {author} {\bibfnamefont {F.}~\bibnamefont
  {Kling}} and \bibinfo {author} {\bibfnamefont {A.}~\bibnamefont {Rajaraman}},
  }\href {\doibase 10.1103/PhysRevD.96.044039} {\bibfield  {journal} {\bibinfo
  {journal} {\emph {Phys. Rev. D}} }\textbf {\bibinfo {volume} {96}}, \bibinfo
  {pages} {044039} (\bibinfo {year} {2017})}, \Eprint
  {http://arxiv.org/abs/1706.04272} {arXiv:1706.04272}\BibitemShut {NoStop}%
\bibitem [{\citenamefont {Annulli} \emph {et~al.}(2020)\citenamefont {Annulli},
  \citenamefont {Cardoso}, and \citenamefont {Vicente}}]{Annulli:2020}%
  \BibitemOpen
  \bibfield  {author} {\bibinfo {author} {\bibfnamefont {L.}~\bibnamefont
  {Annulli}}, \bibinfo {author} {\bibfnamefont {V.}~\bibnamefont {Cardoso}},
  and \bibinfo {author} {\bibfnamefont {R.}~\bibnamefont {Vicente}},
  }\href@noop {} {  (\bibinfo {year} {2020})}\BibitemShut {NoStop}%
\bibitem [{\citenamefont {Guzman} and \citenamefont
  {Urena-Lopez}(2004)}]{Guzman:2004wj}%
  \BibitemOpen
  \bibfield  {author} {\bibinfo {author} {\bibfnamefont {F.S.} \bibnamefont
  {Guzman}} and \bibinfo {author} {\bibfnamefont {L.A.} \bibnamefont
  {Urena-Lopez}}, }\href {\doibase 10.1103/PhysRevD.69.124033} {\bibfield
  {journal} {\bibinfo  {journal} {\emph {Phys. Rev. D}} }\textbf {\bibinfo
  {volume} {69}}, \bibinfo {pages} {124033} (\bibinfo {year} {2004})}, \Eprint
  {http://arxiv.org/abs/gr-qc/0404014} {arXiv:gr-qc/0404014}\BibitemShut
  {NoStop}%
\bibitem [{\citenamefont {Zerilli}(1970)}]{Zerilli:1971wd}%
  \BibitemOpen
  \bibfield  {author} {\bibinfo {author} {\bibfnamefont {F.}~\bibnamefont
  {Zerilli}}, }\href {\doibase 10.1103/PhysRevD.2.2141} {\bibfield  {journal}
  {\bibinfo  {journal} {\emph {Phys.\ Rev.\ D}} }\textbf {\bibinfo {volume}
  {2}}, \bibinfo {pages} {2141} (\bibinfo {year} {1970})}\BibitemShut {NoStop}%
\bibitem [{\citenamefont {Davis} \emph {et~al.}(1971)\citenamefont {Davis},
  \citenamefont {Ruffini}, \citenamefont {Press}, and \citenamefont
  {Price}}]{Davis:1971gg}%
  \BibitemOpen
  \bibfield  {author} {\bibinfo {author} {\bibfnamefont {M.}~\bibnamefont
  {Davis}}, \bibinfo {author} {\bibfnamefont {R.}~\bibnamefont {Ruffini}},
  \bibinfo {author} {\bibfnamefont {W.}~\bibnamefont {Press}},  and \bibinfo
  {author} {\bibfnamefont {R.}~\bibnamefont {Price}}, }\href {\doibase
  10.1103/PhysRevLett.27.1466} {\bibfield  {journal} {\bibinfo  {journal}
  {\emph {Phys.\ Rev.\ Lett.}} }\textbf {\bibinfo {volume} {27}}, \bibinfo
  {pages} {1466} (\bibinfo {year} {1971})}\BibitemShut {NoStop}%
\bibitem [{\citenamefont {Barack} and \citenamefont
  {Pound}(2019)}]{Barack:2018yvs}%
  \BibitemOpen
  \bibfield  {author} {\bibinfo {author} {\bibfnamefont {L.}~\bibnamefont
  {Barack}} and \bibinfo {author} {\bibfnamefont {A.}~\bibnamefont {Pound}},
  }\href {\doibase 10.1088/1361-6633/aae552} {\bibfield  {journal} {\bibinfo
  {journal} {\emph {Rept.\ Prog.\ Phys.}} }\textbf {\bibinfo {volume} {82}},
  \bibinfo {pages} {016904} (\bibinfo {year} {2019})}, \Eprint
  {http://arxiv.org/abs/1805.10385} {arXiv:1805.10385}\BibitemShut {NoStop}%
\bibitem [{\citenamefont {Ben~Menahem} and \citenamefont {Singh}(1982)}]{Ari}%
  \BibitemOpen
  \bibfield  {author} {\bibinfo {author} {\bibfnamefont {A.}~\bibnamefont
  {Ben~Menahem}} and \bibinfo {author} {\bibfnamefont {S.J.} \bibnamefont
  {Singh}}, }\href
  {https://www.amazon.com/Seismic-Waves-Sources-Ari-Ben-Menahem/dp/0486404617}
  {\emph {\bibinfo {title} {Seismic Waves and Sources}}} (\bibinfo  {publisher}
  {Dover publications}, \bibinfo {year} {1982})\BibitemShut {NoStop}%
\bibitem [{\citenamefont {Ostriker}(1999)}]{Ostriker:1998fa}%
  \BibitemOpen
  \bibfield  {author} {\bibinfo {author} {\bibfnamefont {E.C.} \bibnamefont
  {Ostriker}}, }\href {\doibase 10.1086/306858} {\bibfield  {journal} {\bibinfo
   {journal} {\emph {Astrophys. J.}} }\textbf {\bibinfo {volume} {513}},
  \bibinfo {pages} {252} (\bibinfo {year} {1999})}, \Eprint
  {http://arxiv.org/abs/astro-ph/9810324} {arXiv:astro-ph/9810324}\BibitemShut
  {NoStop}%
\bibitem [{\citenamefont {Vicente} \emph {et~al.}(2019)\citenamefont {Vicente},
  \citenamefont {Cardoso}, and \citenamefont {Zilhao}}]{Vicente:2019ilr}%
  \BibitemOpen
  \bibfield  {author} {\bibinfo {author} {\bibfnamefont {R.}~\bibnamefont
  {Vicente}}, \bibinfo {author} {\bibfnamefont {V.}~\bibnamefont {Cardoso}},
  and \bibinfo {author} {\bibfnamefont {M.}~\bibnamefont {Zilhao}}, }\href
  {\doibase 10.1093/mnras/stz2526} {\bibfield  {journal} {\bibinfo  {journal}
  {\emph {Mon.\ Not.\ Roy.\ Astron.\ Soc.}} }\textbf {\bibinfo {volume} {489}},
  \bibinfo {pages} {5424} (\bibinfo {year} {2019})}, \Eprint
  {http://arxiv.org/abs/1905.06353} {arXiv:1905.06353}\BibitemShut {NoStop}%
\bibitem [{\citenamefont {Yoshida} \emph {et~al.}(1994)\citenamefont {Yoshida},
  \citenamefont {Eriguchi}, and \citenamefont {Futamase}}]{Yoshida:1994xi}%
  \BibitemOpen
  \bibfield  {author} {\bibinfo {author} {\bibfnamefont {S.}~\bibnamefont
  {Yoshida}}, \bibinfo {author} {\bibfnamefont {Y.}~\bibnamefont {Eriguchi}},
  and \bibinfo {author} {\bibfnamefont {T.}~\bibnamefont {Futamase}}, }\href
  {\doibase 10.1103/PhysRevD.50.6235} {\bibfield  {journal} {\bibinfo
  {journal} {\emph {Phys.\ Rev.\ D}} }\textbf {\bibinfo {volume} {50}},
  \bibinfo {pages} {6235} (\bibinfo {year} {1994})}\BibitemShut {NoStop}%
\bibitem [{\citenamefont {Kojima} \emph {et~al.}(1991)\citenamefont {Kojima},
  \citenamefont {Yoshida}, and \citenamefont {Futamase}}]{Kojima:1991np}%
  \BibitemOpen
  \bibfield  {author} {\bibinfo {author} {\bibfnamefont {Y.}~\bibnamefont
  {Kojima}}, \bibinfo {author} {\bibfnamefont {S.}~\bibnamefont {Yoshida}},
  and \bibinfo {author} {\bibfnamefont {T.}~\bibnamefont {Futamase}}, }\href
  {\doibase 10.1143/PTP.86.401} {\bibfield  {journal} {\bibinfo  {journal}
  {\emph {Prog.\ Theor.\ Phys.}} }\textbf {\bibinfo {volume} {86}}, \bibinfo
  {pages} {401} (\bibinfo {year} {1991})}\BibitemShut {NoStop}%
\bibitem [{\citenamefont {Macedo} \emph
  {et~al.}(2013{\natexlab{b}})\citenamefont {Macedo}, \citenamefont {Pani},
  \citenamefont {Cardoso}, and \citenamefont {Crispino}}]{Macedo:2013jja}%
  \BibitemOpen
  \bibfield  {author} {\bibinfo {author} {\bibfnamefont {C.F.} \bibnamefont
  {Macedo}}, \bibinfo {author} {\bibfnamefont {P.}~\bibnamefont {Pani}},
  \bibinfo {author} {\bibfnamefont {V.}~\bibnamefont {Cardoso}},  and \bibinfo
  {author} {\bibfnamefont {L.C.B.} \bibnamefont {Crispino}}, }\href {\doibase
  10.1103/PhysRevD.88.064046} {\bibfield  {journal} {\bibinfo  {journal} {\emph
  {Phys.\ Rev.\ D}} }\textbf {\bibinfo {volume} {88}}, \bibinfo {pages}
  {064046} (\bibinfo {year} {2013}{\natexlab{b}})}, \Eprint
  {http://arxiv.org/abs/1307.4812} {arXiv:1307.4812}\BibitemShut {NoStop}%
\bibitem [{\citenamefont {Macedo} \emph {et~al.}(2016)\citenamefont {Macedo},
  \citenamefont {Cardoso}, \citenamefont {Crispino}, and \citenamefont
  {Pani}}]{Macedo:2016wgh}%
  \BibitemOpen
  \bibfield  {author} {\bibinfo {author} {\bibfnamefont {C.F.B.} \bibnamefont
  {Macedo}}, \bibinfo {author} {\bibfnamefont {V.}~\bibnamefont {Cardoso}},
  \bibinfo {author} {\bibfnamefont {L.C.B.} \bibnamefont {Crispino}},  and
  \bibinfo {author} {\bibfnamefont {P.}~\bibnamefont {Pani}}, }\href {\doibase
  10.1103/PhysRevD.93.064053} {\bibfield  {journal} {\bibinfo  {journal} {\emph
  {Phys.\ Rev.\ D}} }\textbf {\bibinfo {volume} {93}}, \bibinfo {pages}
  {064053} (\bibinfo {year} {2016})}, \Eprint {http://arxiv.org/abs/1603.02095}
  {arXiv:1603.02095}\BibitemShut {NoStop}%
\bibitem [{GRI()}]{GRITJHU}%
  \BibitemOpen
  \href@noop {} { }\bibinfo {note}
  {\noindent\url{http://blackholes.ist.utl.pt/?page=Files} \\
  \noindent\url{https://pages.jh.edu/~eberti2/ringdown/}}\BibitemShut {NoStop}%
\bibitem [{\citenamefont {Seidel} and \citenamefont {Suen}(1994)}]{Seidel1994}%
  \BibitemOpen
  \bibfield  {author} {\bibinfo {author} {\bibfnamefont {E.}~\bibnamefont
  {Seidel}} and \bibinfo {author} {\bibfnamefont {W.M.} \bibnamefont {Suen}},
  }\href {\doibase 10.1103/PhysRevLett.72.2516} {\bibfield  {journal} {\bibinfo
   {journal} {\emph {Phys. Rev. Lett.}} }\textbf {\bibinfo {volume} {72}},
  \bibinfo {pages} {2516} (\bibinfo {year} {1994})}\BibitemShut {NoStop}%
\bibitem [{\citenamefont {Mendes} and \citenamefont
  {Yang}(2017)}]{Mendes:2016vdr}%
  \BibitemOpen
  \bibfield  {author} {\bibinfo {author} {\bibfnamefont {R.F.} \bibnamefont
  {Mendes}} and \bibinfo {author} {\bibfnamefont {H.}~\bibnamefont {Yang}},
  }\href {\doibase 10.1088/1361-6382/aa842d} {\bibfield  {journal} {\bibinfo
  {journal} {\emph {Class.\ Quant.\ Grav.}} }\textbf {\bibinfo {volume} {34}},
  \bibinfo {pages} {185001} (\bibinfo {year} {2017})}, \Eprint
  {http://arxiv.org/abs/1606.03035} {arXiv:1606.03035}\BibitemShut {NoStop}%
\bibitem [{\citenamefont {Cardoso} \emph {et~al.}(2017)\citenamefont {Cardoso},
  \citenamefont {Franzin}, \citenamefont {Maselli}, \citenamefont {Pani}, and
  \citenamefont {Raposo}}]{Cardoso:2017cfl}%
  \BibitemOpen
  \bibfield  {author} {\bibinfo {author} {\bibfnamefont {V.}~\bibnamefont
  {Cardoso}}, \bibinfo {author} {\bibfnamefont {E.}~\bibnamefont {Franzin}},
  \bibinfo {author} {\bibfnamefont {A.}~\bibnamefont {Maselli}}, \bibinfo
  {author} {\bibfnamefont {P.}~\bibnamefont {Pani}},  and \bibinfo {author}
  {\bibfnamefont {G.}~\bibnamefont {Raposo}}, }\href {\doibase
  10.1103/PhysRevD.95.084014} {\bibfield  {journal} {\bibinfo  {journal} {\emph
  {Phys.\ Rev.\ D}} }\textbf {\bibinfo {volume} {95}}, \bibinfo {pages}
  {084014} (\bibinfo {year} {2017})}, \bibinfo {note} {[Addendum: Phys.Rev.D
  95, 089901 (2017)]}, \Eprint {http://arxiv.org/abs/1701.01116}
  {arXiv:1701.01116}\BibitemShut {NoStop}%
\bibitem [{\citenamefont {Sennett} \emph {et~al.}(2017)\citenamefont {Sennett},
  \citenamefont {Hinderer}, \citenamefont {Steinhoff}, \citenamefont
  {Buonanno}, and \citenamefont {Ossokine}}]{Sennett:2017etc}%
  \BibitemOpen
  \bibfield  {author} {\bibinfo {author} {\bibfnamefont {N.}~\bibnamefont
  {Sennett}}, \bibinfo {author} {\bibfnamefont {T.}~\bibnamefont {Hinderer}},
  \bibinfo {author} {\bibfnamefont {J.}~\bibnamefont {Steinhoff}}, \bibinfo
  {author} {\bibfnamefont {A.}~\bibnamefont {Buonanno}},  and \bibinfo {author}
  {\bibfnamefont {S.}~\bibnamefont {Ossokine}}, }\href {\doibase
  10.1103/PhysRevD.96.024002} {\bibfield  {journal} {\bibinfo  {journal} {\emph
  {Phys.\ Rev.\ D}} }\textbf {\bibinfo {volume} {96}}, \bibinfo {pages}
  {024002} (\bibinfo {year} {2017})}, \Eprint {http://arxiv.org/abs/1704.08651}
  {arXiv:1704.08651}\BibitemShut {NoStop}%
\bibitem [{\citenamefont {Gondolo} and \citenamefont
  {Silk}(1999)}]{Gondolo:1999ef}%
  \BibitemOpen
  \bibfield  {author} {\bibinfo {author} {\bibfnamefont {P.}~\bibnamefont
  {Gondolo}} and \bibinfo {author} {\bibfnamefont {J.}~\bibnamefont {Silk}},
  }\href {\doibase 10.1103/PhysRevLett.83.1719} {\bibfield  {journal} {\bibinfo
   {journal} {\emph {Phys. Rev. Lett.}} }\textbf {\bibinfo {volume} {83}},
  \bibinfo {pages} {1719} (\bibinfo {year} {1999})}, \Eprint
  {http://arxiv.org/abs/astro-ph/9906391} {arXiv:astro-ph/9906391}\BibitemShut
  {NoStop}%
\bibitem [{\citenamefont {Sadeghian} \emph {et~al.}(2013)\citenamefont
  {Sadeghian}, \citenamefont {Ferrer}, and \citenamefont
  {Will}}]{Sadeghian:2013laa}%
  \BibitemOpen
  \bibfield  {author} {\bibinfo {author} {\bibfnamefont {L.}~\bibnamefont
  {Sadeghian}}, \bibinfo {author} {\bibfnamefont {F.}~\bibnamefont {Ferrer}},
  and \bibinfo {author} {\bibfnamefont {C.M.} \bibnamefont {Will}}, }\href
  {\doibase 10.1103/PhysRevD.88.063522} {\bibfield  {journal} {\bibinfo
  {journal} {\emph {Phys. Rev. D}} }\textbf {\bibinfo {volume} {88}}, \bibinfo
  {pages} {063522} (\bibinfo {year} {2013})}, \Eprint
  {http://arxiv.org/abs/1305.2619} {arXiv:1305.2619}\BibitemShut {NoStop}%
\bibitem [{\citenamefont {Merritt} \emph {et~al.}(2002)\citenamefont {Merritt},
  \citenamefont {Milosavljevic}, \citenamefont {Verde}, and \citenamefont
  {Jimenez}}]{Merritt:2002vj}%
  \BibitemOpen
  \bibfield  {author} {\bibinfo {author} {\bibfnamefont {D.}~\bibnamefont
  {Merritt}}, \bibinfo {author} {\bibfnamefont {M.}~\bibnamefont
  {Milosavljevic}}, \bibinfo {author} {\bibfnamefont {L.}~\bibnamefont
  {Verde}},  and \bibinfo {author} {\bibfnamefont {R.}~\bibnamefont {Jimenez}},
  }\href {\doibase 10.1103/PhysRevLett.88.191301} {\bibfield  {journal}
  {\bibinfo  {journal} {\emph {Phys. Rev. Lett.}} }\textbf {\bibinfo {volume}
  {88}}, \bibinfo {pages} {191301} (\bibinfo {year} {2002})}, \Eprint
  {http://arxiv.org/abs/astro-ph/0201376} {arXiv:astro-ph/0201376}\BibitemShut
  {NoStop}%
\bibitem [{\citenamefont {Bertone} and \citenamefont
  {Merritt}(2005)}]{Bertone:2005hw}%
  \BibitemOpen
  \bibfield  {author} {\bibinfo {author} {\bibfnamefont {G.}~\bibnamefont
  {Bertone}} and \bibinfo {author} {\bibfnamefont {D.}~\bibnamefont {Merritt}},
  }\href {\doibase 10.1103/PhysRevD.72.103502} {\bibfield  {journal} {\bibinfo
  {journal} {\emph {Phys. Rev. D}} }\textbf {\bibinfo {volume} {72}}, \bibinfo
  {pages} {103502} (\bibinfo {year} {2005})}, \Eprint
  {http://arxiv.org/abs/astro-ph/0501555} {arXiv:astro-ph/0501555}\BibitemShut
  {NoStop}%
\bibitem [{\citenamefont {Merritt}(2004)}]{Merritt:2003qk}%
  \BibitemOpen
  \bibfield  {author} {\bibinfo {author} {\bibfnamefont {D.}~\bibnamefont
  {Merritt}}, }\href {\doibase 10.1103/PhysRevLett.92.201304} {\bibfield
  {journal} {\bibinfo  {journal} {\emph {Phys. Rev. Lett.}} }\textbf {\bibinfo
  {volume} {92}}, \bibinfo {pages} {201304} (\bibinfo {year} {2004})}, \Eprint
  {http://arxiv.org/abs/astro-ph/0311594} {arXiv:astro-ph/0311594}\BibitemShut
  {NoStop}%
\bibitem [{\citenamefont {Herdeiro} and \citenamefont
  {Radu}(2015)}]{Herdeiro:2015waa}%
  \BibitemOpen
  \bibfield  {author} {\bibinfo {author} {\bibfnamefont {C.A.} \bibnamefont
  {Herdeiro}} and \bibinfo {author} {\bibfnamefont {E.}~\bibnamefont {Radu}},
  }\href {\doibase 10.1142/S0218271815420146} {\bibfield  {journal} {\bibinfo
  {journal} {\emph {Int. J. Mod. Phys. D}} }\textbf {\bibinfo {volume} {24}},
  \bibinfo {pages} {1542014} (\bibinfo {year} {2015})}, \Eprint
  {http://arxiv.org/abs/1504.08209} {arXiv:1504.08209}\BibitemShut {NoStop}%
\bibitem [{\citenamefont {Cardoso} and \citenamefont
  {Gualtieri}(2016)}]{Cardoso:2016ryw}%
  \BibitemOpen
  \bibfield  {author} {\bibinfo {author} {\bibfnamefont {V.}~\bibnamefont
  {Cardoso}} and \bibinfo {author} {\bibfnamefont {L.}~\bibnamefont
  {Gualtieri}}, }\href {\doibase 10.1088/0264-9381/33/17/174001} {\bibfield
  {journal} {\bibinfo  {journal} {\emph {Class. Quant. Grav.}} }\textbf
  {\bibinfo {volume} {33}}, \bibinfo {pages} {174001} (\bibinfo {year}
  {2016})}, \Eprint {http://arxiv.org/abs/1607.03133}
  {arXiv:1607.03133}\BibitemShut {NoStop}%
\bibitem [{Note1()}]{Note1}%
  \BibitemOpen
  \bibinfo {note} {Actually, such a sphere should be placed outside the
  effective potential for wave propagation around BHs, but the distinction is
  not relevant here.}\BibitemShut {Stop}%
\bibitem [{\citenamefont {Unruh}(1976)}]{Unruh:1976fm}%
  \BibitemOpen
  \bibfield  {author} {\bibinfo {author} {\bibfnamefont {W.}~\bibnamefont
  {Unruh}}, }\href {\doibase 10.1103/PhysRevD.14.3251} {\bibfield  {journal}
  {\bibinfo  {journal} {\emph {Phys. Rev. D}} }\textbf {\bibinfo {volume}
  {14}}, \bibinfo {pages} {3251} (\bibinfo {year} {1976})}\BibitemShut
  {NoStop}%
\bibitem [{\citenamefont {Herdeiro} and \citenamefont
  {Radu}(2014)}]{Herdeiro:2014goa}%
  \BibitemOpen
  \bibfield  {author} {\bibinfo {author} {\bibfnamefont {C.A.R.} \bibnamefont
  {Herdeiro}} and \bibinfo {author} {\bibfnamefont {E.}~\bibnamefont {Radu}},
  }\href {\doibase 10.1103/PhysRevLett.112.221101} {\bibfield  {journal}
  {\bibinfo  {journal} {\emph {Phys.Rev.Lett.}} }\textbf {\bibinfo {volume}
  {112}}, \bibinfo {pages} {221101} (\bibinfo {year} {2014})}, \Eprint
  {http://arxiv.org/abs/1403.2757} {arXiv:1403.2757}\BibitemShut {NoStop}%
\bibitem [{\citenamefont {Brito} \emph {et~al.}(2015)\citenamefont {Brito},
  \citenamefont {Cardoso}, and \citenamefont {Pani}}]{Brito:2015oca}%
  \BibitemOpen
  \bibfield  {author} {\bibinfo {author} {\bibfnamefont {R.}~\bibnamefont
  {Brito}}, \bibinfo {author} {\bibfnamefont {V.}~\bibnamefont {Cardoso}},  and
  \bibinfo {author} {\bibfnamefont {P.}~\bibnamefont {Pani}}, }\href {\doibase
  10.1007/978-3-319-19000-6} {\bibfield  {journal} {\bibinfo  {journal} {\emph
  {Lect. Notes Phys.}} }\textbf {\bibinfo {volume} {906}}, \bibinfo {pages}
  {pp.1} (\bibinfo {year} {2015})}, \Eprint {http://arxiv.org/abs/1501.06570}
  {arXiv:1501.06570}\BibitemShut {NoStop}%
\bibitem [{\citenamefont {{Bekenstein}}(1973)}]{1973ApJ...183..657B}%
  \BibitemOpen
  \bibfield  {author} {\bibinfo {author} {\bibfnamefont {J.D.} \bibnamefont
  {{Bekenstein}}}, }\href {\doibase 10.1086/152255} {\bibfield  {journal}
  {\bibinfo  {journal} {\emph {\apj}} }\textbf {\bibinfo {volume} {183}},
  \bibinfo {pages} {657} (\bibinfo {year} {1973})}\BibitemShut {NoStop}%
\bibitem [{\citenamefont {Lancaster} \emph {et~al.}(2020)\citenamefont
  {Lancaster}, \citenamefont {Giovanetti}, \citenamefont {Mocz}, \citenamefont
  {Kahn}, \citenamefont {Lisanti}, and \citenamefont
  {Spergel}}]{Lancaster:2019mde}%
  \BibitemOpen
  \bibfield  {author} {\bibinfo {author} {\bibfnamefont {L.}~\bibnamefont
  {Lancaster}}, \bibinfo {author} {\bibfnamefont {C.}~\bibnamefont
  {Giovanetti}}, \bibinfo {author} {\bibfnamefont {P.}~\bibnamefont {Mocz}},
  \bibinfo {author} {\bibfnamefont {Y.}~\bibnamefont {Kahn}}, \bibinfo {author}
  {\bibfnamefont {M.}~\bibnamefont {Lisanti}},  and \bibinfo {author}
  {\bibfnamefont {D.N.} \bibnamefont {Spergel}}, }\href {\doibase
  10.1088/1475-7516/2020/01/001} {\bibfield  {journal} {\bibinfo  {journal}
  {\emph {JCAP}} }\textbf {\bibinfo {volume} {01}}, \bibinfo {pages} {001}
  (\bibinfo {year} {2020})}, \Eprint {http://arxiv.org/abs/1909.06381}
  {arXiv:1909.06381}\BibitemShut {NoStop}%
\bibitem [{\citenamefont {Gualandris} and \citenamefont
  {Merritt}(2008)}]{Gualandris:2007nm}%
  \BibitemOpen
  \bibfield  {author} {\bibinfo {author} {\bibfnamefont {A.}~\bibnamefont
  {Gualandris}} and \bibinfo {author} {\bibfnamefont {D.}~\bibnamefont
  {Merritt}}, }\href {\doibase 10.1086/586877} {\bibfield  {journal} {\bibinfo
  {journal} {\emph {Astrophys. J.}} }\textbf {\bibinfo {volume} {678}},
  \bibinfo {pages} {780} (\bibinfo {year} {2008})}, \Eprint
  {http://arxiv.org/abs/0708.0771} {arXiv:0708.0771}\BibitemShut {NoStop}%
\bibitem [{\citenamefont {Graham} \emph {et~al.}(2020)}]{Graham:2020gwr}%
  \BibitemOpen
  \bibfield  {author} {\bibinfo {author} {\bibfnamefont {M.}~\bibnamefont
  {Graham}} \emph {et~al.}, }\href {\doibase 10.1103/PhysRevLett.124.251102}
  {\bibfield  {journal} {\bibinfo  {journal} {\emph {Phys. Rev. Lett.}}
  }\textbf {\bibinfo {volume} {124}}, \bibinfo {pages} {251102} (\bibinfo
  {year} {2020})}, \Eprint {http://arxiv.org/abs/2006.14122}
  {arXiv:2006.14122}\BibitemShut {NoStop}%
\bibitem [{\citenamefont {Flanagan} and \citenamefont
  {Hughes}(1998)}]{Flanagan:1997sx}%
  \BibitemOpen
  \bibfield  {author} {\bibinfo {author} {\bibfnamefont {E.E.} \bibnamefont
  {Flanagan}} and \bibinfo {author} {\bibfnamefont {S.A.} \bibnamefont
  {Hughes}}, }\href {\doibase 10.1103/PhysRevD.57.4535} {\bibfield  {journal}
  {\bibinfo  {journal} {\emph {Phys. Rev. D}} }\textbf {\bibinfo {volume}
  {57}}, \bibinfo {pages} {4535} (\bibinfo {year} {1998})}, \Eprint
  {http://arxiv.org/abs/gr-qc/9701039} {arXiv:gr-qc/9701039}\BibitemShut
  {NoStop}%
\bibitem [{\citenamefont {Coleman}(1985)}]{Coleman:1985ki}%
  \BibitemOpen
  \bibfield  {author} {\bibinfo {author} {\bibfnamefont {S.R.} \bibnamefont
  {Coleman}}, }\href {\doibase 10.1016/0550-3213(86)90520-1} {\bibfield
  {journal} {\bibinfo  {journal} {\emph {Nucl. Phys. B}} }\textbf {\bibinfo
  {volume} {262}}, \bibinfo {pages} {263} (\bibinfo {year} {1985})}, \bibinfo
  {note} {[Erratum: Nucl.Phys.B 269, 744 (1986)]}\BibitemShut {NoStop}%
\end{thebibliography}%
\end{document}